\documentclass[aps,prd,reprint,nofootinbib,showkeys,showpacs,superscriptaddress,longbibliography,floatfix]{revtex4-2}

\pdfoutput=1

\usepackage{amsmath}
\usepackage{amssymb}
\usepackage[american]{babel}
\usepackage{booktabs}
\usepackage{dcolumn}  
\usepackage[T1]{fontenc}  
\usepackage{graphicx}  
\usepackage[colorlinks]{hyperref}  
\usepackage[utf8]{inputenc}  
\usepackage{multirow}
\usepackage{txfonts}  
\usepackage{url}
\usepackage[dvipsnames]{xcolor}
\usepackage{float}
\usepackage{lipsum}
\newcolumntype{d}[1]{D{.}{.}{#1}}
\newcolumntype{v}[1]{D{,}{,\ }{#1}}
\makeatletter

\newcommand{\Rmnum}[1]{\expandafter\@slowromancap\romannumeral #1@}
\makeatother

\hypersetup{
	citecolor = blue,
	linkcolor = RoyalPurple,
	urlcolor = blue
}

\begin{document}

\title{Cosmological perturbations in the Tsallis holographic dark energy scenarios}

\author{W. J. C. da Silva}
\email{williamjouse@fisica.ufrn.br}
\affiliation{Universidade Federal do Rio Grande do Norte, Departamento de F\'{\i}sica, Natal - RN, 59072-970, Brazil}

\author{R. Silva}
\email{raimundosilva@fisica.ufrn.br}
\affiliation{Universidade Federal do Rio Grande do Norte, Departamento de F\'{\i}sica, Natal - RN, 59072-970, Brazil}
\affiliation{Universidade do Estado do Rio Grande do Norte, Departamento de F\'{\i}sica, Mossor\'o - RN, 59610-210, Brazil}

\date{\today}

\begin{abstract}
We investigate the Tsallis holographic dark energy (THDE) models in the context of perturbations' growth. We assume the description of dark energy by considering the holographic principle and the nonadditive entropy to carry out this. We implement the perturbed relativistic equations to achieve the growth of matter fluctuations, being the growth rate of the cosmic structures is non-negligible at low redshifts. To constrain and compare the models, we carry out the Bayesian analysis using the recent geometrical and growth rate observational data. The main results are: (i) the models are compatible with cosmological observations, (ii) the cosmological constant recovered with a $1\sigma$ confidence level, furthermore (iii) they could cross the phantom barrier.  Finally, the models can relieve $\approx 1\sigma$ the $\sigma_8$ tension in the non-clustered case and can alleviate in $\approx 2.8\sigma$ the $H_0$ tension. From the model selection viewpoint, the data discarded the THDE models.

\end{abstract}

\maketitle

\section{Introduction}\label{}

The universe's accelerated expansion is based on several cosmological observable \cite{Riess1999,Perlmutter1999,Weinberg2012,ade2015, aghanim2018}. These cosmological probes converge to the so-called standard model - the $\Lambda$CDM, i.e., a cosmological constant $\Lambda$ plus Cold Dark Matter (CDM). Although the success, there are both theoretical and observational issues associated with the cosmological constant problem, physics the dark sector, general relativity as a gravity theory on large scales, anomalies in the cosmic microwave background and, the tensions in the measurements of Hubble constant and matter fluctuations (For instance, see Refs. \cite{weinberg1988, Padmanabhan2002, riess2016, riess2018, Macaulay2013, Bull2015, Freedman2017, DelPopolo2016}). There are diverse ways to relieve these issues. Frequently ones can be divided into modified general relativity
\cite{Clifton2011} and dark energy models \cite{Huterer2017}. The first situation adopts modifications in the general relativity, and a new description of dark energy (fluid or scalar field) has been considered in the dark energy models.

From the theoretical standpoint, exist an alternative to address DE origin, e.g. the holographic principle of quantum gravity \cite{tHooft1993, Susskind1994} at a cosmological framework \cite{Fischler1998,Bak1999,Horava2000}. In this approach, the hypothesis on the relation between the ultraviolet cutoff and the entropy of a system is used \cite{Cohen1999}, being that a class of models has been possible,i.e., the holographic dark energy (HDE) \cite{Li2004, Wang2016}. In the flat Friedmann-Lema\^itre-Robertson-Walker (FLRW) universe, the HDE used the Bekenstein-Hawking entropy for the cosmological horizon and considered Hubble horizon as its infrared cutoff \cite{Hsu2004, Li2004, Guberina2006}. Indeed, the HDE related to the Hubble horizon is inconsistent due to the dark energy density provides a similar evolution to matter density \cite{Hsu2004, Li2004}. Nevertheless, the HDE presenting different cutoffs leads to other cosmological approaches \cite{Li2004, Huang2004, Pavon2005, Wang2005, Nojiri2005, Wang2016, Wang2016b, Nojiri2017, Malekjani2018, Dai:2020rfo}. 

The nonadditive entropy, on the other hand, is the core of the nonextensive framework \cite{Tsallis1988, Gell2004}, which has been used in the cosmological context (see Refs. \cite{Komatsu2013,Komatsu2014,Komatsu2014-2,Nunes2015,Tavayef2018,Saridakis2018,Tsallis2012,Gimenes2018,dasilva2019,Silva2019,Sadri2019}, and the references therein). Some HDE models have been proposed \cite{Jahromi2018, Moradpour2018} to introduce the nonadditive entropy in the HDE context. In particular, these approaches recover the Bekenstein entropy in the additive limit \cite{Majhi2017, Abe2001, Touchette2002}. The study that connects the HDE and nonadditive entropy used a black hole entropy-area relation \cite{Tsallis2012}. By applying this modification and taking the holographic principle into account, the authors of Ref. \cite{Tavayef2018} proposed the holographic dark energy. This model was unable to obtain the standard HDE as a limit case, where the Hubble horizon as cutoff provided an incorrect description compared with the standard HDE \cite{Li2004, Hsu2004, Jahromi2018, Moradpour2018}. Afterward, considering different cutoffs, some THDE models were proposed to improve the connection between Tsallis entropy and HDE \cite{Saridakis2018,DAgostino2019,Sadri2019,Zadeh2018,Zadeh2018-2,Ghaffari2018,Nojiri2019}.

In the perturbative level, there are effects on the formation of large-scale structures of the universe associated with the DE, e.g., the dynamical DE models
\cite{Abramo2007, Abramo2009, Tsujikawa2013, Batista2014, Mehrabi2015, Rezaei2017, Mehrabi2018}.  The redshift-space distortions (RSD) provide measurement to obtain DE models' effect upon clustering of the matter \cite{Abramo2007, Tsujikawa2013, Batista2014, Mehrabi2015, Rezaei2017, Mehrabi2018}. Some works consider the perturbation level to investigate the HDE models \cite{Kao2005, Xu2013, Mehrabi2015, DAgostino2019}. 

Here, we will investigate the cosmological constraints in the HDE models through the perturbative level and RSD data. To do this, we assume three different HDE models, i.e., standard HDE with future event horizon as cutoff, Tsallis HDE as Hubble horizon, and future event horizon as cutoffs \cite{Li2004,Tavayef2018,Sadri2019,Saridakis2018,DAgostino2019}, we investigate these models in the perturbative level. Specifically, we implement the cosmological perturbations equations to examine the influence of the HDE on the growth of matter. In order to distinguish the holographic models, we perform a Bayesian model comparison analysis using the most recent baryon acoustic oscillations measurements (BAO), measurement on the baryons from Big Bang nucleosynthesis (BBN), cosmic microwave background priors (CMB), cosmic chronometers (CC), supernovae type Ia from Pantheon sample, and redshift-space distortions data.

The paper is organized as follows. In Sect. \ref{sec:THDE} we present the holographic dark energy models considered in this work. In Sect. \ref{sec:perturbations}, we resume the cosmological perturbations equations and show the influence of the HDE models on the perturbation framework. In Sect. \ref{sec:data-results}, we show the cosmological data and the Bayesian model comparison framework. In Sect. \ref{sec:results}, we implement this analysis to get limits on the cosmological parameters, and we study these models from the viewpoint of both the evidence and the Bayes factor. Finally, in Sect. \ref{sec:conclusions} we summarize the results.

\section{Holographic dark energy model}\label{sec:THDE}
The holographic principle says that all information enclosed by a region of space can be described as a hologram \cite{tHooft1993, Susskind1994}. Especially, as a consequence of the holographic principle, the number of degrees of freedom from a finite-size system should be finite, as well as bounded through the corresponding area of its boundary \cite{Cohen1999}. Then, if $\rho_{\Lambda}$ represents the vacuum energy density caused by a short distance cutoff  (ultraviolet (UV) cutoff), the total energy in a physical region of size $L$ (infrared (IR) cutoff) should be smaller than the mass of a black hole of equal size, thus the following relation is satisfied $L^3\rho_{\Lambda} \le LM^{2}_{P\ell}$ ($M^{2}_{P\ell}$ is the Planck mass). Based on these arguments, Li proposed the so-called holographic dark energy (HDE) model \citep{Li2004}. The holographic dark energy density is given by
\begin{equation}\label{eq:hde-density}
	\rho_{\text{HDE}} =\frac{3C^2M^{2}_{P\ell}}{L^2},
\end{equation}
where often, for simplicity, the $C$ parameter is assumed constant \cite{Li2004,Malekjani2018}. It is crucial to choose the characteristic length scale of $L$. Indeed, several models of holographic dark energy are generated by different length scales of $L$ \cite{Wang2016}.

Using a different black hole entropy-area relationship, proposed in Ref. \cite{Tsallis2012}, together with the holographic principle, a new holographic dark energy was introduced by \citet{Tavayef2018}. This model assumes a flat FLRW universe based on both definition and derivation from the standard holographic energy density. The holographic dark energy density depends on the entropy-area relationship of black holes through $S \sim A \sim L^2$, being $A = 4\pi L^2$ the area of the horizon \cite{Cohen1999}. The modification proposed by Tavayef et al. \cite{Tavayef2018} is associated with the horizon entropy of a black hole given by \cite{Tsallis2012}
\begin{equation}\label{entropy}
S_{\delta} = \gamma A^\delta,
\end{equation}
with $\gamma$ and $\delta$ being unknown constants. These parameters under the hypothesis of equal probabilities are related to the dimensionality of the system $d$, in particular situation is $\delta = d/(d-1)$ for $d > 1$ \cite{Tsallis2012}, however in the general case they remain as free parameters. The Bekenstein entropy is a particular case when $\delta \rightarrow 1$ and $\gamma =2\pi M^{2}_{P\ell}$(in units where $\hbar = k_{\rm{B}} = c = 1$). Moreover, the power-law distribution of probability converges for the usual one \cite{Tsallis2012}. Equivalent result was calculated through the quantum gravity \cite{Rashki2014} and has been used to study cosmological and holographic frameworks \cite{Komatsu2013, Komatsu2014, Komatsu2014-2, Moradpour2016}.

Considering the holographic hypothesis, Cohen et al. proposed the relation between the system entropy, $S$, infrared ($L$) and ultraviolet ($\Lambda$) cutoffs given by
\begin{equation}
	L^3 \Lambda^3  \le S^{\frac{3}{4}},
\end{equation}
which from Eq. (\ref{entropy}) one finds
\begin{equation}
\Lambda^4 \le (4\pi)^{\delta}\gamma L^{2\delta - 4},
\end{equation}
being $\Lambda^4$ the vacuum energy density \cite{Li2004, Guberina2006, Wang2016, Akhlaghi2018, Malekjani2018}. Then, from this inequality, recently was proposed the new HDE density, i.e., the Tsallis holographic dark energy (THDE) \cite{Tavayef2018}
\begin{equation}\label{densityTHDE}
	\rho_{\text{THDE}} = BL^{2\delta - 4},
\end{equation}
where $B = 3C^2M^{2}_{P\ell}$. In particular, for  $\delta = 1$, the standard HDE is recovered, Eq. (\ref{eq:hde-density}). Furthermore, for $\delta = 2$, the cosmological constant model is retrieved.  The important point is that the THDE density also is defined in terms of the IR cutoff $L$. More recently, some approaches have been proposed using different scales of the $L$ \cite{Saridakis2018, Zadeh2018, Zadeh2018-2}. We use here two characteristic scales: the Hubble horizon, the future event horizon. Moreover, the standard HDE with future event horizon as a characteristic scale.

Let us summarize below the holographic energy models investigated through the Bayesian analysis. 

\subsection{THDE with Hubble horizon cutoff}

The simplest choice for scale $L$ is the Hubble horizon, i.e., $L = H^{-1}$. The authors in \cite{Tavayef2018} proposed this case, where they obtained that this description can lead to the late time accelerated universe, even without considering the interaction between the components of the dark sector. The result obtained is in opposition to the standard HDE model with Hubble horizon cutoff, which does not imply the accelerated expansion, except the interaction, is taken into account. That way, from Eq. (\ref{densityTHDE}), it is possible to find out the THDE density \cite{Tavayef2018}
\begin{equation}\label{density-tsallis}
\rho_{\rm de} = BH^{-2\delta + 4},
\end{equation}
where $B$ is an unknown parameter and $H$ is the Hubble parameter.

In the framework of a flat FLRW universe filled with radiation, pressure-less matter, and THDE, the Friedmann equation reads
\begin{equation}\label{friedmann-eq}
	H^{2} = \frac{1}{3M^{2}_{P\ell}}(\rho_{\text{r}} + \rho_{\text{m}} + \rho_{\text{de}}),
\end{equation}
where $\rho_{\text{r}}$, $\rho_{\text{m}}$ and $\rho_{\text{de}}$ are the radiation, matter (cold dark matter and baryon) and, dark energy densities, respectively. In the case of a non-interacting system for which the cosmic components evolve separately, it is possible to write the conservation laws equations which describe the density evolution of each cosmic fluid as
\begin{eqnarray}
&\dot{\rho}_{\rm r}+4H\rho_{\rm r}=0, \label{conservation-radiation} \\
&\dot{\rho}_{\rm m}+3H\rho_{\rm m}=0, \label{conservation-baryon} \\
&\dot{\rho}_{\rm de}+3H(1+w_{\rm de})\rho_{\rm de}=0,  \label{conservation-dark}
\end{eqnarray}
where the over dot is the derivative with respect to time and $\omega_{\rm de}$ is the equation of state parameter of holographic dark energy.  

Considering the time derivative of Eq. (\ref{friedmann-eq}), using Eqs. (\ref{density-tsallis}), (\ref{conservation-radiation}), (\ref{conservation-baryon}) and, (\ref{conservation-dark}) it leads to \cite{Tavayef2018}
\begin{equation}\label{eos-thde}
	w_{\rm de}(x) = \frac{\delta - 1}{(2-\delta)\Omega_{\rm de} - 1},
\end{equation}
where $\delta$ is the nonadditive parameter, $\Omega_{\rm de}$ is the dimensionless density parameter of the THDE. In the case where $\delta < 1$, it follows that $2 - \delta > 1$ and there is a divergence in the evolution of $w_{\rm de}$ arisen at the redshift for which $\Omega_{\rm de} = \frac{1}{2-\delta}$. Hence, $\delta < 1$  is not compatible with this model. According to Eq. (\ref{eos-thde}), we can find that in the early universe $w_{\rm de} \rightarrow 1 - \delta$ since $\Omega_{\rm de} \rightarrow 0$ and in the future $w_{\rm de} \rightarrow - 1$ since $\Omega_{\rm de} \rightarrow 1$. 

Now, taking the time derivative of $\Omega_{\rm de} = (B/3M^{2}_{P\ell})H^{-2\delta + 2}$, using Eq. (\ref{eos-thde}), it is possible to show the equation governing the dynamical evolution of the THDE model in terms of $x = \ln a = -\ln(1+z)$ \cite{Tavayef2018}
\begin{equation}\label{edo-THDE}
	\frac{d\Omega_{\rm de}}{dx} = 3(\delta - 1)\Omega_{\rm de}\left(\frac{1-\Omega_{\rm de}}{1-(2-\delta)\Omega_{\rm de}}\right).
\end{equation}
Since $0 < \Omega_{\rm de} < 1$, $\frac{d\Omega_{\rm de}}{dz}$ is always negative, specifically the density of THDE always increases with redshift $z \rightarrow -1$. Indeed, this implies that the universe's expansion will never-changing point so that the universe will not re-collapse in the future.

Now, using the Friedmann Eq. (\ref{friedmann-eq}) and the conservation Eqs. (\ref{conservation-radiation}), (\ref{conservation-baryon}) and (\ref{conservation-dark}), the Hubble parameter of the THDE model is written as
\begin{equation}\label{H-THDE}
	\frac{H^{2}(z)}{H^{2}_{0}} = \frac{\Omega_{\text{r}}(1+z)^{4} + \Omega_{\text{m}}(1+z)^{3}}{1 - \Omega_{\rm de}(z)},
\end{equation}
where $\Omega_{\text{r}}$ and, $\Omega_{\text{m}}$ represent the current values of the radiation and matter density parameters, respectively. Solving numerically Eq. (\ref{edo-THDE}) and substituting the interpolated results into Eq. (\ref{H-THDE}), the redshift evolution of Hubble parameter $H(z)$ of the THDE model should be obtained.

\subsection{THDE with future event horizon cutoff}

Another choice for characteristic scale $L$ is the future event horizon
\begin{equation}\label{event-horizont}
	R_{h} = a \int_{t}^{\infty} \frac{dt}{a} = a\int_{a}^{\infty}\frac{da}{a^2H},
\end{equation}
where $a$ is the scale factor, $t$ is the cosmic time and $H$, Hubble parameter. This model was recently investigated in Refs. \cite{Saridakis2018, DAgostino2019}, where they argue that the model in which the scale $L = H^{-1}$ (THDE, here) presented a disadvantage due it does not recover the HDE as a limit case (see Refs. \cite{Saridakis2018, DAgostino2019}, and the references therein).

By considering that $L = R_{h}$, the energy density of THDE Eq. (\ref{density-tsallis}) is given by
\begin{equation}\label{eq:THDE2-density}
	\rho_{\rm de} = BR_{h}^{2\delta - 4},
\end{equation}
where $B$ and $\delta$ are constants. Notice for $\delta = 2$ the cosmological constant $\rho_{\rm de} = \Lambda$ is recovered. Now, using the definition $\Omega_{\rm de} = (B/3M^{2}_{P\ell})R_{h}^{2\delta - 4}$ and Eq. (\ref{event-horizont}), one finds
\begin{equation}\label{relation-THDE2}
	a \int_{a}^{\infty} \frac{da}{a^2H} = \left(\frac{B}{3M^{2}_{P\ell}H^2\Omega_{\rm de}}\right)^{\frac{1}{4-2\delta}}.
\end{equation}
Writing Eq. (\ref{H-THDE}) in terms of the scale factor and inserting into Eq. (\ref{relation-THDE2}), it follows
\begin{equation}\label{relation2-THDE2}
	\int_{x}^{\infty} \frac{dx}{H_{0}\sqrt{\Omega_{\text{m}}}}\sqrt{a(1 - \Omega_{\rm de})}=\frac{1}{a} \left(\frac{B}{3M^{2}_{P\ell}H^2\Omega_{\rm de}}\right)^{\frac{1}{4-2\delta}},
\end{equation}
where $x = \ln a = -\ln(1+z)$ as the independent variable. Differentiating (\ref{relation2-THDE2}) with respect to $x$ one obtains
\begin{equation}\label{EDO-THDE2}
	\frac{1}{\Omega_{\rm de}(1 - \Omega_{\rm de})}\frac{d\Omega_{\rm de}}{dx} = 2\delta - 1 + Q(1-\Omega_{\rm de})^{\frac{1-\delta}{2(2-\delta)}}\Omega_{\rm de}^{\frac{1}{2(2-\delta)}}e^{\frac{3(1-\delta)}{2(2-\delta)}x},
\end{equation}
with
\begin{equation}
	Q = 2(2-\delta)\left(\frac{B}{3M^{2}_{P\ell}}\right)^\frac{1-\delta}{2(2-\delta)}\left(H_{0}\sqrt{\Omega_{\text{m}}}\right)^{\frac{1}{2(2-\delta)}}.
\end{equation}
Eq. (\ref{EDO-THDE2}) is the differential equation in which the solution determines the evolution of the Tsallis holographic dark energy density. In the case where $\delta = 1$ this model behaves like standard holographic dark energy \cite{Li2004}.

The equation of state parameter $w_{\rm de}$ can be obtained from Eqs. (\ref{event-horizont}), (\ref{eq:THDE2-density}), (\ref{H-THDE}) and (\ref{conservation-dark}), it is given by
\begin{equation}\label{eq:eos-thde2}
	w_{\rm de}(x) = \frac{1}{3} - \frac{2\delta}{3} - \frac{Q}{3}\Omega_{\rm de}^{\frac{1}{2(2-\delta)}}(1-\Omega_{\rm de}^{\frac{\delta -1}{2(2-\delta)}})e^{\frac{3(1-\delta)}{2(2-\delta)}x}.
\end{equation}
This equation provides the evolution of equation of state parameter as a function of $x$, since is possible to know the solution of Eq. (\ref{EDO-THDE2}). Note that for $\delta = 1$ and identifying $B = 3C^2M^{2}_{P\ell}$, the equation of state parameter of the standard holographic dark energy Eq. (\ref{eq:conservation-hde}), should be recovered. According the Ref. \cite{Saridakis2018}, for increasing $\delta$, $w_{\rm de}$ evolution tends to lower values. For $\delta \geq 1.2$, $w_{\rm de}(z = 0)$ assumes the value compatible with phantom regime. We named this model as THDE2.

\subsection{Standard HDE}
For completeness we considered the standard HDE proposed in the Ref. \cite{Li2004}. This model is characterized by future event horizon as length scale $L$, Eq. (\ref{event-horizont}). Then, from dark energy conservation equation (\ref{conservation-dark}) and taking the time derivative of Eq. (\ref{eq:hde-density}) with $L = R_{h}$, it is possible to obtain equation of state parameter \cite{Li2004}
\begin{equation}\label{eq:conservation-hde}
	w_{\rm de}(x) = -\frac{1}{3} -\frac{2}{3C}\sqrt{\Omega_{\rm de}},
\end{equation}
where $\Omega_{\rm de}$ is the density parameter of the standard HDE. Now, taking derivative of (\ref{eq:hde-density}) with respect to redshift, using Eq. (\ref{event-horizont}) and Eq. (\ref{eq:conservation-hde}),
it follows the differential equation governing the dynamical evolution of HDE model
\begin{equation}\label{eq:hde-evolution}
	\frac{d\Omega_{\rm de}}{dx} = \Omega_{\rm de}(1 - \Omega_{\rm de})\Bigg(1 + \frac{2\sqrt{\Omega_{\rm de}}}{C}\Bigg).
\end{equation}
Once that $0 < \Omega_{\rm de} < 1$, $\frac{d\Omega_{\rm de}}{dz}$ is always negative, namely the density of HDE increases along with redshift $z \rightarrow -1$. From Friedmann Eqs. (\ref{friedmann-eq}) and (\ref{eq:hde-evolution}),  the Hubble parameter for this model reads

\begin{equation}\label{eq:H-HDE}
\frac{H^{2}(z)}{H^{2}_{0}} = \frac{\Omega_{\text{r}}(1+z)^{4} + \Omega_{\text{m}}(1+z)^{3}}{1 - \Omega_{\rm de}(z)}.
\end{equation}
The parameter $C$ does an important role in order to determine the cosmic evolution of dark energy in this model. Indeed, if $C = 1$, $w_{\rm de} = -1$ thus HDE will be behave as cosmological constant $\Lambda$; for $C > 1$, $w_{\rm de} > - 1$, thus HDE will tend a quintessence dark energy scenario. Finally, $C < 1$, $w_{\rm de} < -1$, the HDE can cross the phantom line. This means that $C$ is the key parameter that determines the property of HDE.

\section{Cosmological perturbations evolution}\label{sec:perturbations}

\begin{figure*}[t]
	\centering
	\includegraphics[width=8.5cm]{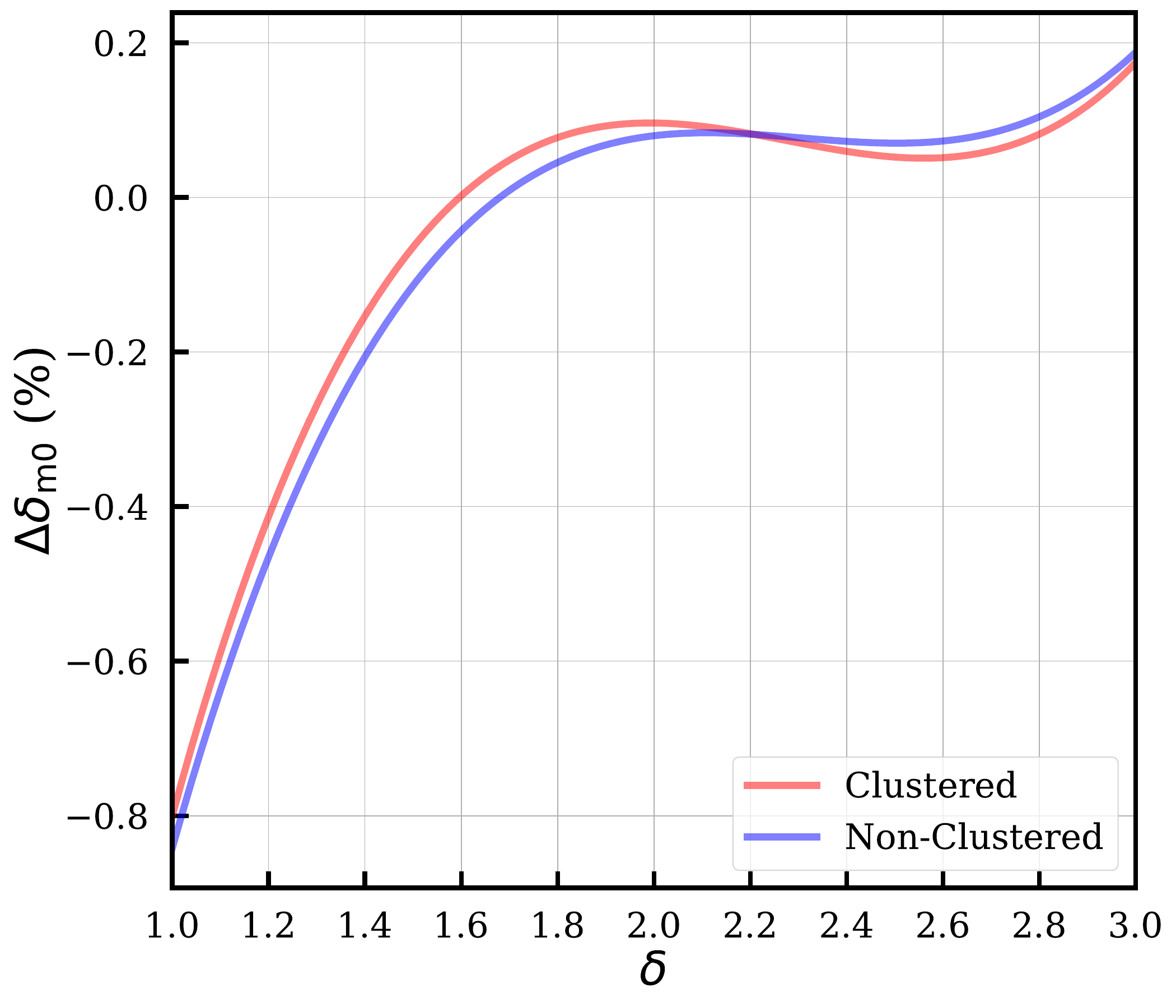}
	\includegraphics[width=8.5cm]{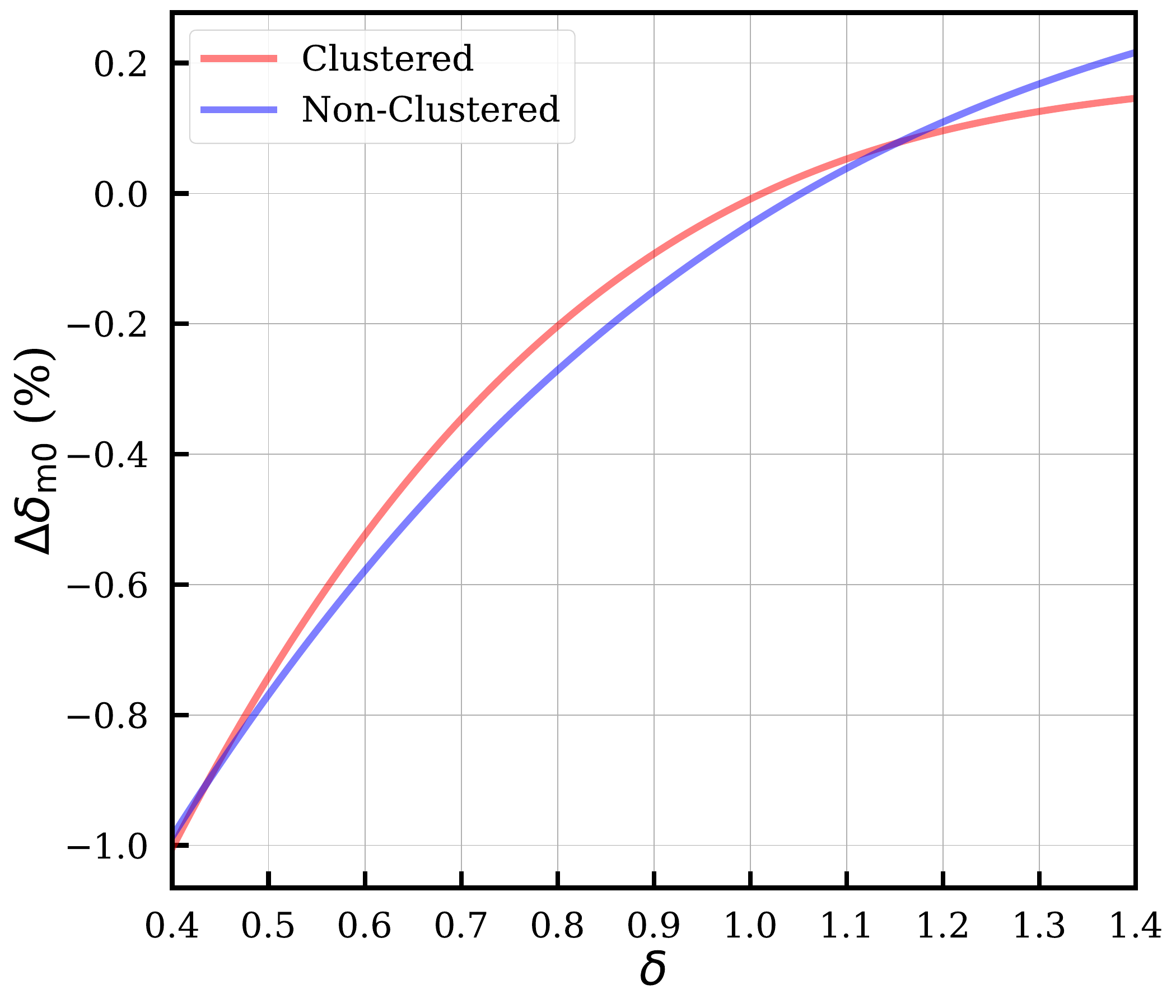}
	\includegraphics[width=8.5cm]{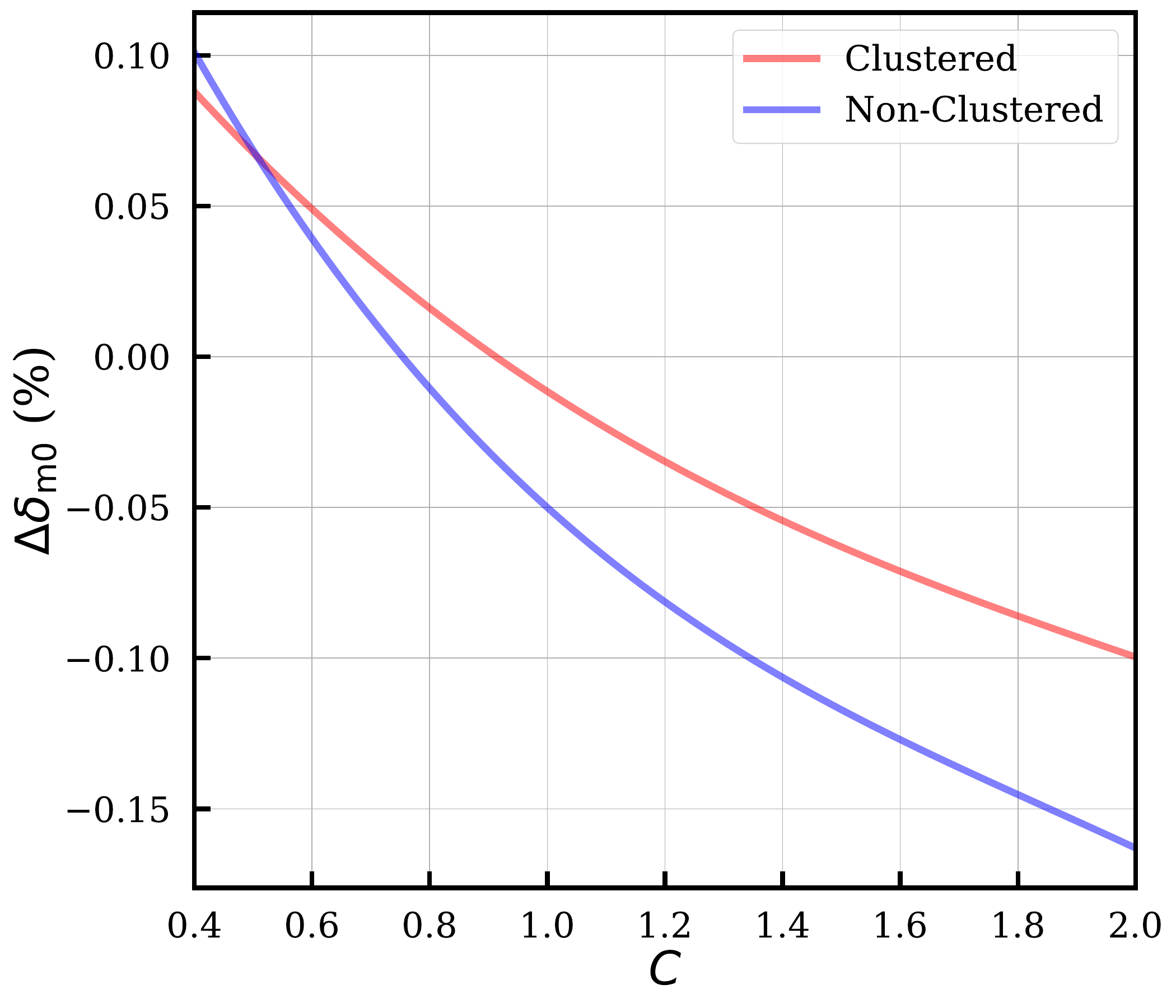}
	\caption{Relative deviation $\Delta \delta_{\rm m0} = (\delta_{\rm m0} - \delta_{\Lambda\rm{CDM}})/\delta_{\Lambda\rm{CDM}}$ of the THDE models matter fluctuations computed at the current time with respect to $\Lambda$CDM as a function of model parameter ($\delta$ and $C$). The upper left is THDE model, upper right is the THDE2 model with $B = 3$ (in units where $M^{2}_{P\ell} = 1$), and lower center is HDE.}
	\label{fig:Delta_d_m}
\end{figure*}

Here, let us recall the main equations of the linear perturbation theory considering HDE models. 
For a complete description see Refs. \cite{Ma1995,Dodelson:2003ft,Piattella2018}. Based on the approach investigated in Ref. \cite{Abramo2009,Mehrabi2015}, let us consider the scalar fluctuations of the FLRW metric in the conformal Newtonian gauge given by
\begin{equation}
ds^2 =  a^2(t)[(1+2\phi)d\eta^2 -(1-2\phi)\delta_{ij}dx^{i}dx^{j}],
\end{equation}
where $\eta$, $a$ and $\phi$ are the conformal time, the scale factor and the Bardeen potential, respectively. The Einstein equations in the perturbed FLRW metric are given by
\begin{eqnarray}
3\mathcal{H}\phi^{\prime}&+&\left(3\mathcal{H}^2+k^2\right)\phi =  
-\frac{3\mathcal{H}^2}{2}(\Omega_{\rm m}\delta_{\rm m}+\Omega_{\rm de}\delta_{\rm de})\;,
\label{eq:first-order-En-eq1}\\
\phi^{\prime\prime}&+&3\mathcal{H}\phi^{\prime}+
\left(\frac{2a^{\prime \prime}}{a}-\mathcal{H}^2\right)\phi  = 
\frac{3\mathcal{H}^2}{2}\Omega_{\rm de}c^{2}_{\rm eff}\delta_{\rm de}\;,
\label{eq:first-order-En-eq2}
\end{eqnarray}
with $\mathcal{H} = aH$ being the Hubble parameter as a function of the conformal time, $c^{2}_{\rm eff}=\frac{\delta p_{\rm de}}{\delta\rho_{\rm de}}$ is the effective sound speed and, the prime is associated to conformal time derivative. In these equations, $\delta_{\rm m}$ and $\delta_{\rm de}$ are the dark matter and dark energy perturbations, respectively. Observe that in Eqs. (\ref{eq:first-order-En-eq1}) and (\ref{eq:first-order-En-eq2}), we consider a general case in which both dark matter and dark energy have been perturbed. For sub-horizon scales ($\mathcal{H}^2\ll k^2$) and in matter domination epoch ($\phi\approx cte$), Eq. (\ref{eq:first-order-En-eq1}) turns on the Poisson equation
\begin{equation}\label{eq:pos}
k^2\phi = -\frac{3}{2}\mathcal{H}^2(\Omega_{\rm m}\delta_{\rm m}+\Omega_{\rm de}\delta_{\rm de}).
\end{equation}

The continuity equations at the linear perturbation level for general fluid are
\begin{eqnarray}
\delta^{\prime}_{i} & = & -(1+w_i)(\theta_i-3\phi^{\prime})-
3\mathcal{H}\left(c^{2}_{\rm eff}-w_i\right)\delta_i,\label{eq:first-order-conser1}\\
\theta^{\prime}_{i} & = & -\mathcal{H}(1-3w_i)\theta_i-\frac{w_i^{\prime}}{1+w_i}\theta_i+
\frac{c^{2}_{\rm eff}}{1+w_i}k^2\delta+k^2\phi,\label{eq:first-order-conser2}
\end{eqnarray}
where $\theta$ is divergence of velocity. The quantity of HDE clustering depends on the magnitude of its effective sound speed $c^{2}_{\rm eff}$. In the case of $c^{2}_{\rm eff} = 0$ HDE clusters in a similar way to dark matter. Nonetheless, due to the existence of the dark energy pressure one may suppose that the magnitude of the HDE perturbations is low with relation to dark matter. In this work we consider HDE as a perfect fluid and we set $c^{2}_{\rm eff} = 0$, which implies that the effective sound 
corresponds with the adiabatic sound speed \cite{Bean2004}
\begin{equation}
	c^2_{\rm a} = w_{\rm de} - \frac{a\frac{d{w}_{\rm de}}{da}}{3(1+w_{\rm de})},
\end{equation}
which is determined by the equation of state parameter $w_{\rm de}$ and so is negative for a lot of dark energy models. 

We can eliminating $\theta$ from the system of equations (\ref{eq:first-order-conser1}) and (\ref{eq:first-order-conser2}), and using $\frac{d}{d\eta} = a^2H\frac{d}{da}$ we obtain, after some manipulations, the following second-order differential equations which describe the evolution of matter and DE perturbations, respectively,
\begin{eqnarray}
\label{eq:sec-ord-delta_m}\frac{d^{2}\delta_{\rm m}}{da^{2}}&+&  A_{\rm m}\frac{d\delta_{\rm m}}{da} + B_{\rm m}\delta_{\rm m} \\  \nonumber &=& \frac{3}{2a^2}(\Omega_{\rm m}\delta_{\rm m} + \Omega_{\rm de}\delta_{\rm de}), \\
\label{eq:sec-ord-delta_d}
\frac{d^{2}\delta_{\rm de}}{da^{2}}&+& A_{\rm de}\frac{d\delta_{\rm de}}{da} + B_{\rm de}\delta_{\rm de} \\  \nonumber &=& \frac{3}{2a^2}(1+w_{\rm de})(\Omega_{\rm m}\delta_{\rm m} + \Omega_{\rm de}\delta_{\rm de}), 
\end{eqnarray}
where the coefficients are given by
\begin{eqnarray}
A_{\rm m} & = & \frac{3}{2a}(1 - w_{\rm de}\Omega_{\rm de}),   \nonumber\\
B_{\rm m} & = & 0,   \nonumber\\     
A_{\rm de} & = & \frac{1}{a}\left[-3w_{\rm de} - \frac{a}{1+w_{\rm de}}\frac{dw_{\rm de}}{da} + \frac{3}{2} (1-w_{\rm de}\Omega_{\rm de})\right], \nonumber \\ 
B_{\rm de} & = & \frac{1}{a^2}\left[-a \frac{dw_{\rm de}}{da} + \frac{aw_{\rm de}}{1+w_{\rm de}}\frac{dw_{\rm de}}{da} - \frac{1}{2}w_{\rm de}(1-3w_{\rm de}\Omega_{\rm de})\right] \nonumber.
\end{eqnarray}

To solve the system of equations (\ref{eq:sec-ord-delta_m}) and (\ref{eq:sec-ord-delta_d}) from $a = 0.001	$ to $a = 1$, we consider the following initial conditions 

\begin{eqnarray}\label{eq:ini}
\frac{d\delta_{\rm m,i}}{da}&=&\frac{\delta_{\rm m,i}}{a_{\rm i}},\\
\delta_{\rm de,i}&=&\frac{1+w_{\rm de,i}}{1-3w_{\rm de,i}}\delta_{\rm m,i},\\
\frac{d\delta_{\rm de,i}}{da}&=&\frac{4\delta_{\rm m,i}}{(1-3w_{\rm de,i })^2}\frac{dw_{\rm de,i}}{da} + \frac{1+w_{\rm de,i}}{1-3w_{\rm de,i}}\frac{d\delta_{\rm m,i}}{da}, 
\end{eqnarray}
where we set $\delta_{\rm m,i} = 1.5 \times 10^{-4}$ which ensures that matter perturbations are in the linear regime ($\delta_{\rm m} \ll 1$). In these conditions, $w_{\rm de,i}$ is computed in $a_{\rm i}$. In this work, we concentrate on the two cases for holographic dark energy models: (i) the dark energy models endure homogeneous ($\delta_{\rm de} = 0$) and the only corresponding matter component clusters (non-clustered); (ii) the HDE models and the matter clustering, namely, $c_{\rm eff} = 0$ (clustered).

\begin{figure*}[t]
	\centering
	\includegraphics[width=8.5cm]{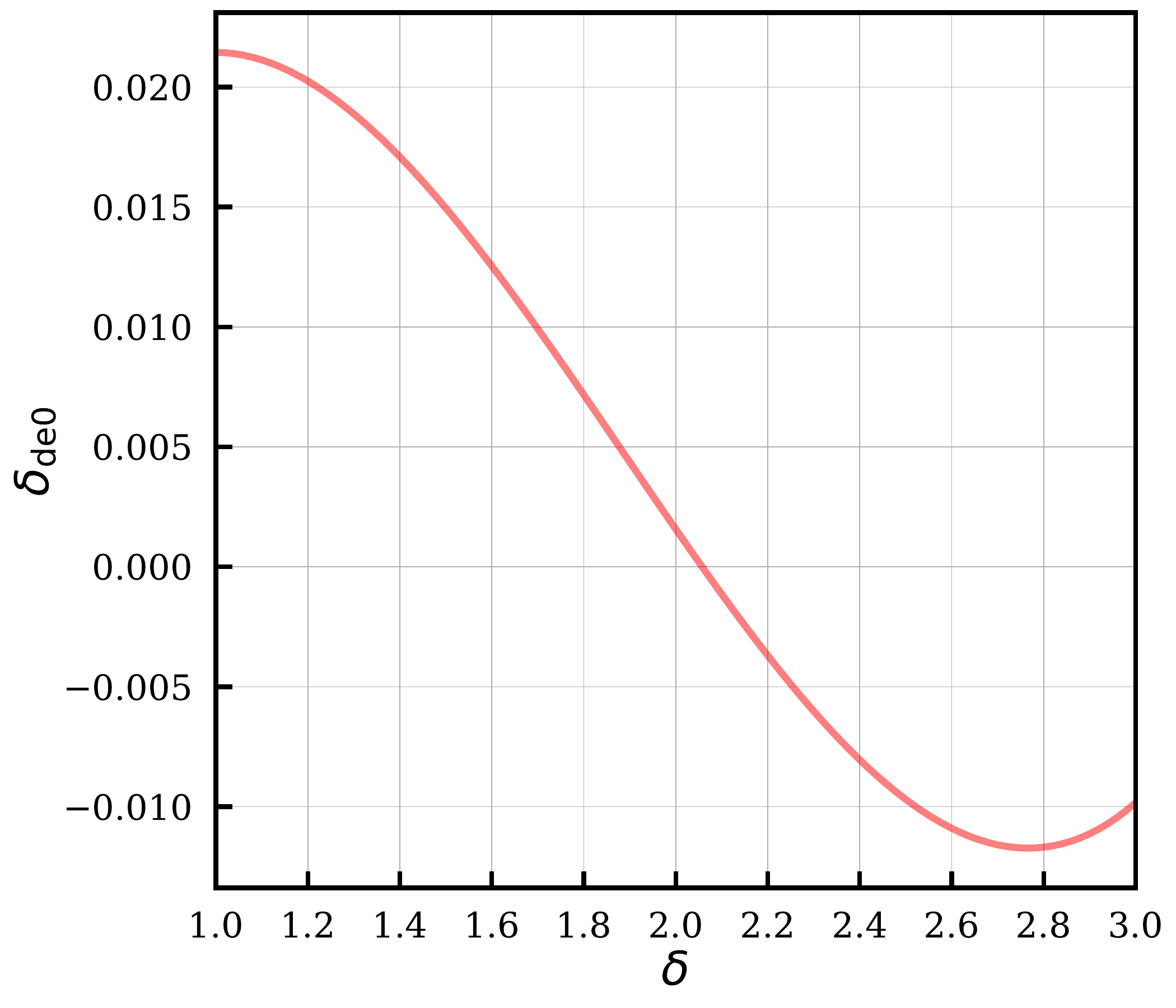}
	\includegraphics[width=8.5cm]{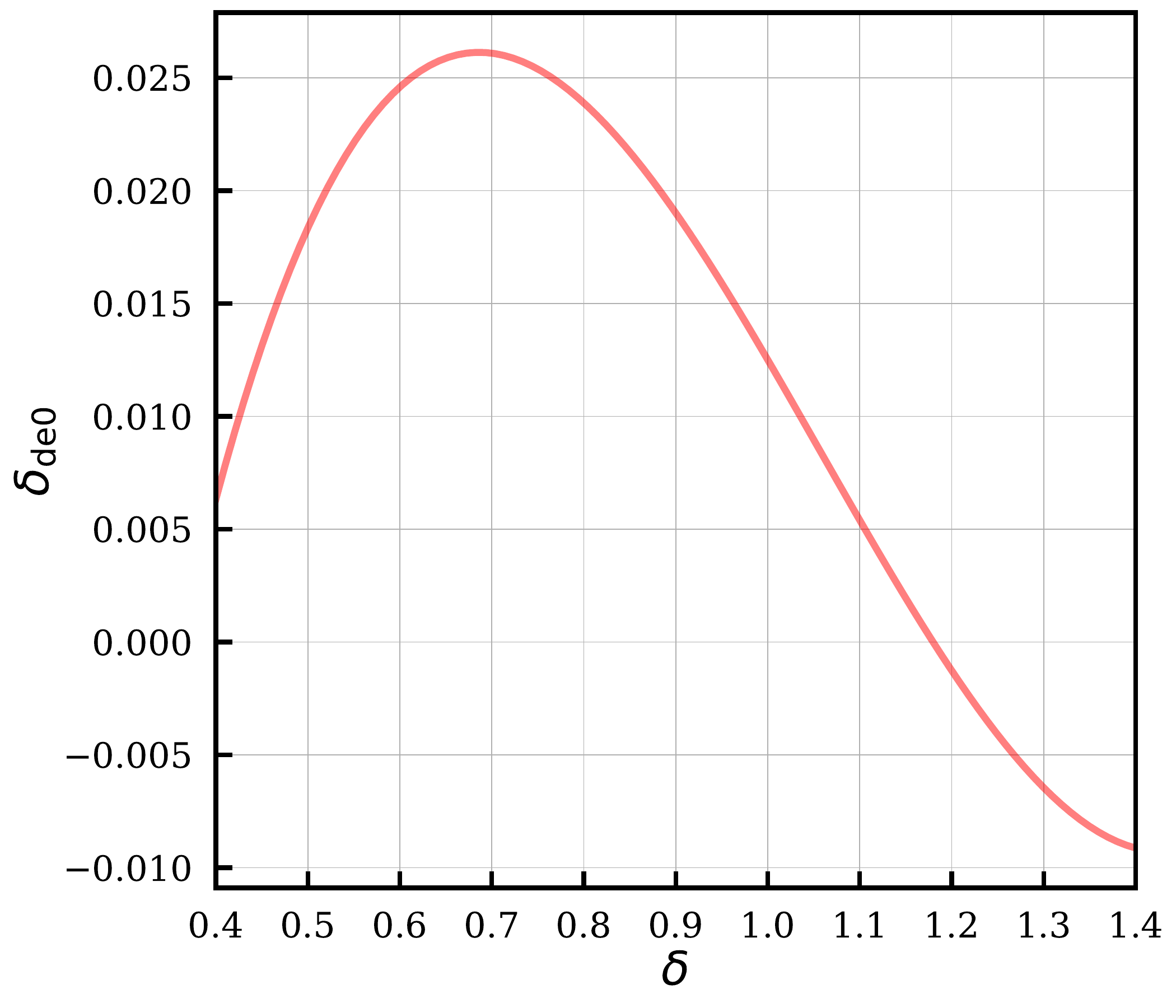}
	\includegraphics[width=8.5cm]{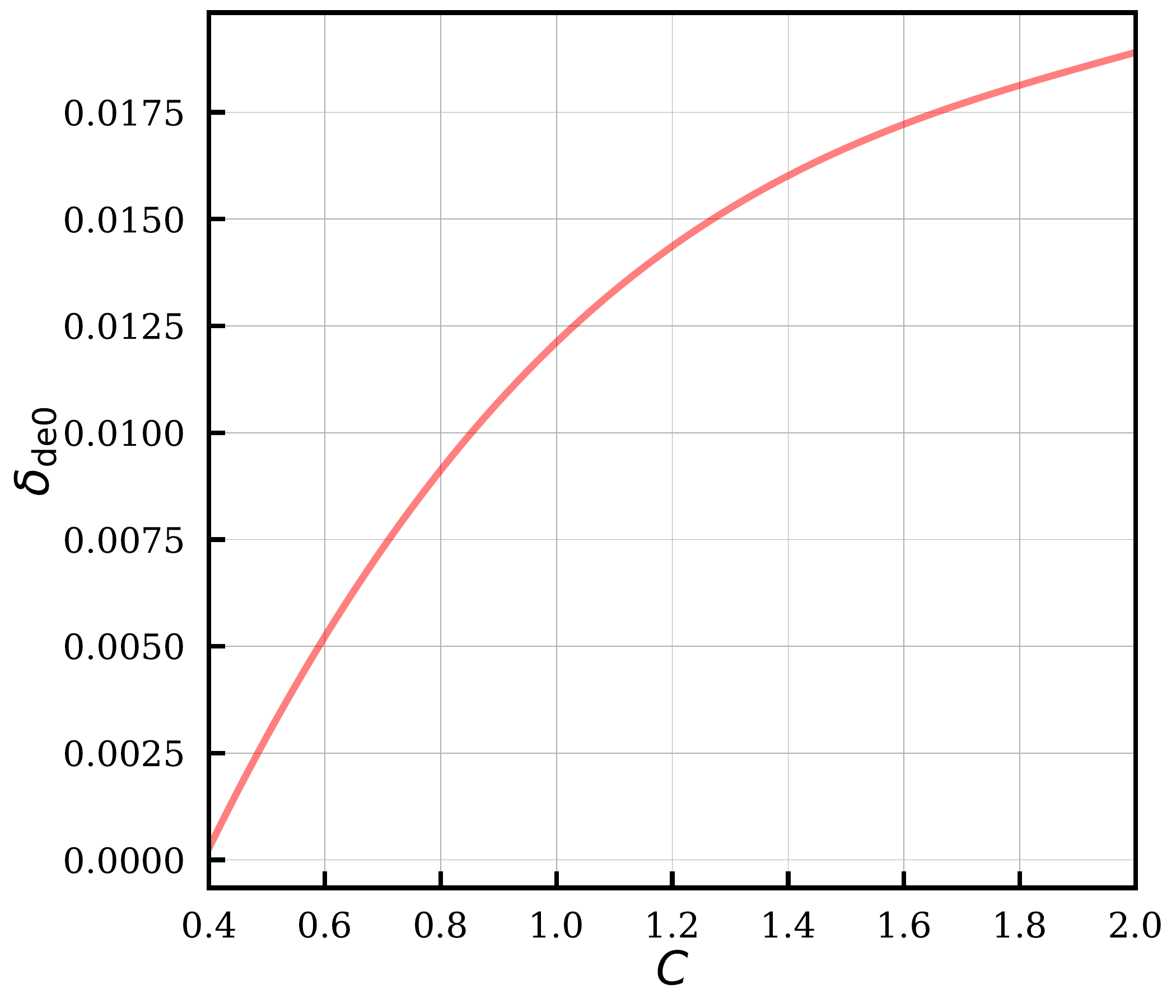}
	\caption{\label{fig:delta_de}THDE contrast computed in the present time as a function of the parameter models. The upper left is THDE model, upper right is THDE2 model with $B = 3$ (in units where $M^{2}_{P\ell} = 1$), and lower center is HDE.}
	
\end{figure*}

\begin{figure*}[t]
	\centering
	\includegraphics[width=8.5cm]{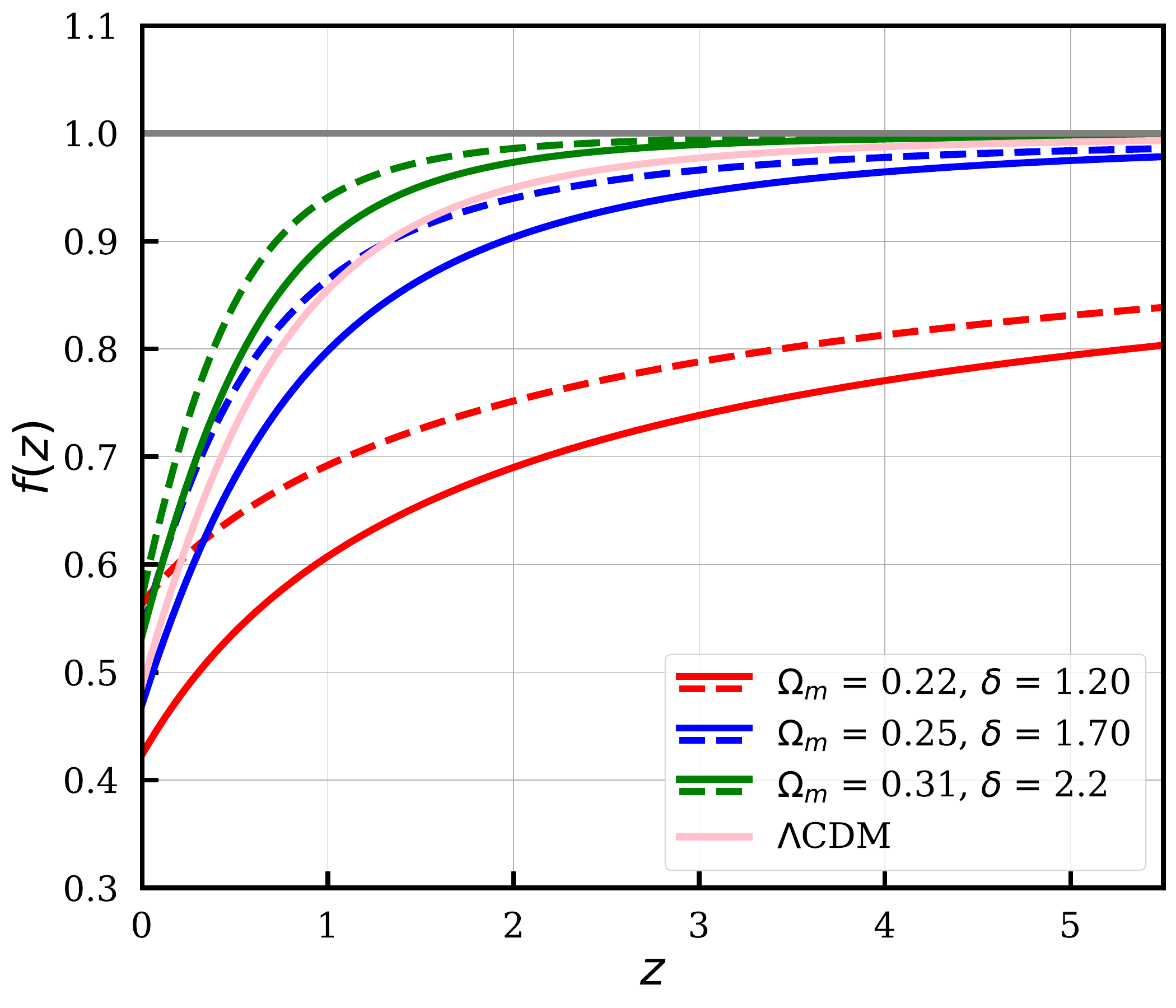}
	\includegraphics[width=8.5cm]{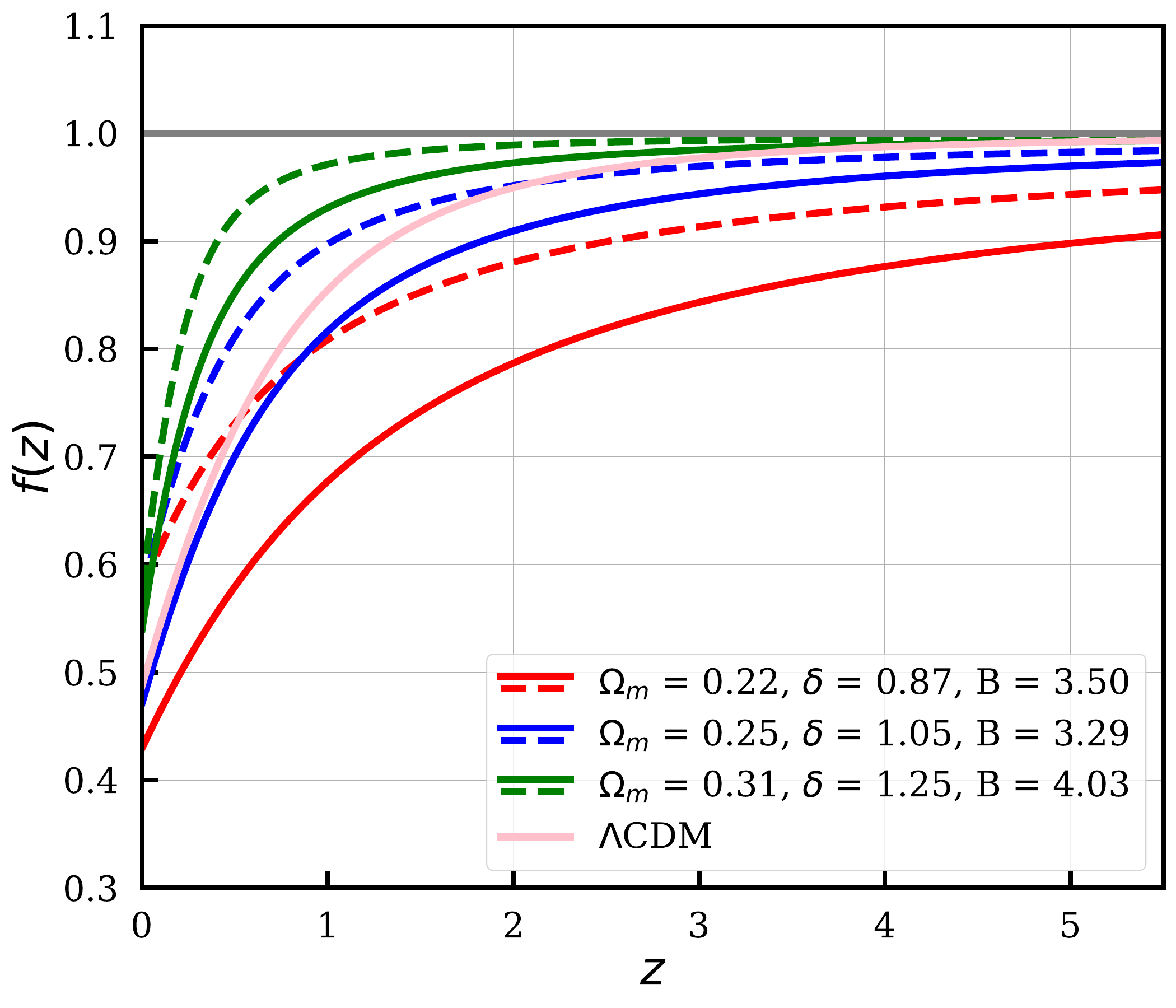}
	\includegraphics[width=8.5cm]{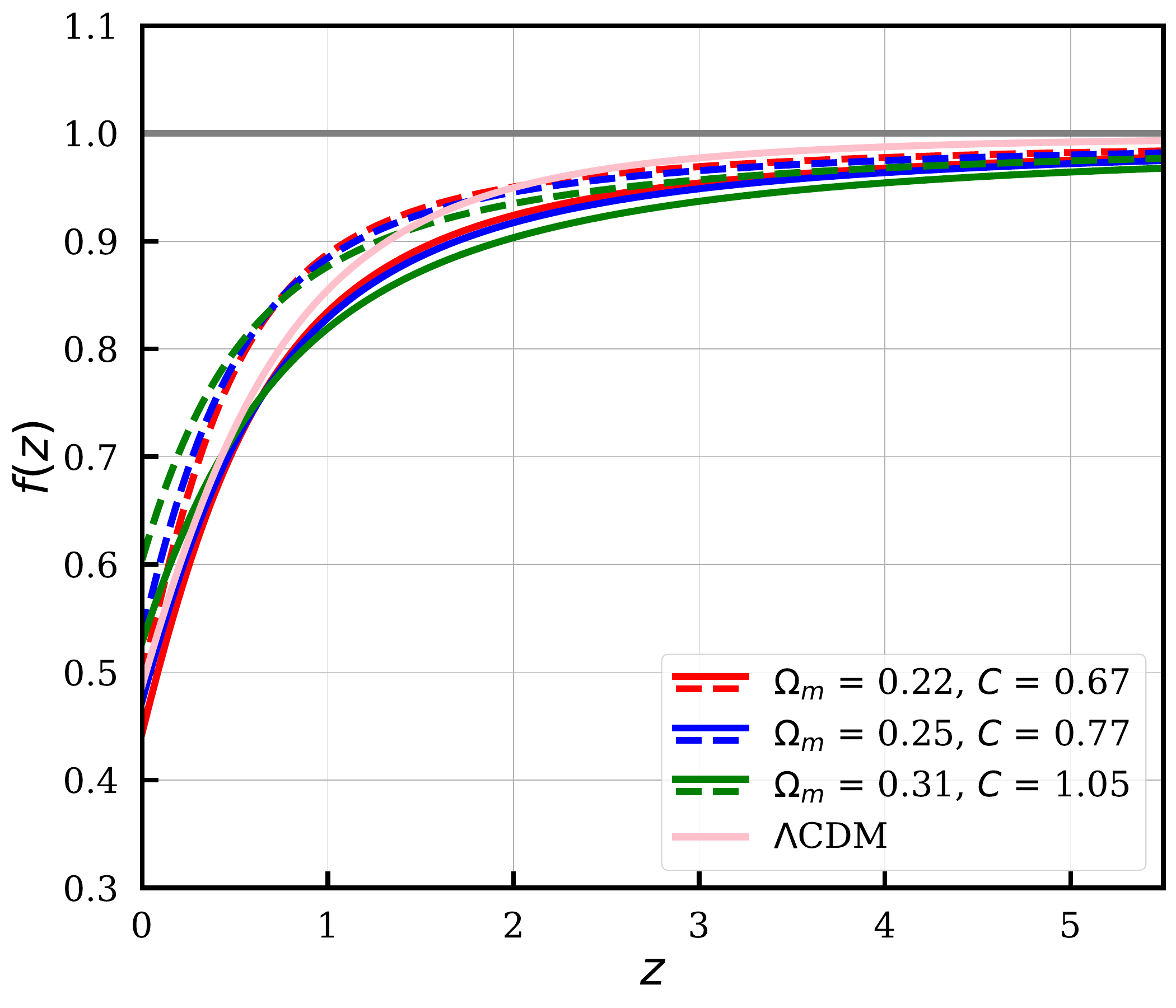}
	\caption{\label{fig:growth-rate}Linear growth rate for different parameter combinations for the clustered and non-clustered THDE in comparison of $\Lambda$CDM. The continuous line is the evolution of  non-clustered HDE and dashed line is the clustered. The upper left is THDE model, upper right is the THDE2 model, and lower center is HDE.}
\end{figure*}

The differential equations Eqs. (\ref{eq:sec-ord-delta_m}) and  (\ref{eq:sec-ord-delta_d}) has a solution with decaying and growing modes. We are concerned about the growing modes that gave the structures that are observed today in the universe. One can then define the linear growth function of perturbations $ D_{\rm m}(a)$ relating the perturbation $\delta_{\rm m}(a)$ at some given scale factor to its value at initial scale factor $a_{\rm i}$

\begin{equation}\label{eq:growth-function}
\delta_{\rm m}(a) = D_{\rm m}(a)\delta_{\rm m}(a = 1).
\end{equation}
This relation is normalized in $D_{\rm m}(a = 1) = 1$. This relation shows how much the perturbations have grown since initial moment $a_{\rm i}$

A significant quantity in investigating the growth of large scale structure is the linear growth function given by

\begin{equation}\label{eq:linear-growth-rate}
f(a) = \frac{d\ln  D_{\rm m}}{d\ln a}.
\end{equation}

A more powerful and reliable quantity that is measured by redshift surveys is the combination of the linear growth rate $f(a)$ and the root-mean-square mass fluctuation in spheres with radius $8\text{h}^{-1}\text{Mpc}$, $\sigma_8(a)$. In the linear regime, one has \cite{Nesseris2008,Song2009,Huterer2015,Ishak2018}

\begin{equation}\label{eq:sigma8}
\sigma_8(z) = \frac{\delta_{\rm m}(z)}{\delta_{\rm m}(z = 0)}\sigma_8(z = 0),
\end{equation}
and,

\begin{equation}\label{eq:fs8}
f\sigma_8(z) = -(1+z)\frac{\sigma_8(z = 0)}{\delta_{\rm m}(z = 0)}\frac{d \delta_{\rm m}}{dz}.
\end{equation}

In practice, most of the $f\sigma_8$ measurements are made via the peculiar velocities obtained from Redshift-Space Distortions (RSD) measures coming from galaxy redshift surveys \cite{Kaiser1987}. As we will discuss further, we will use perturbation formalism described in this section and RSD data to constrain the parameters of the models studied in this work.

In Fig. \ref{fig:Delta_d_m}, we show the relative deviation of the THDE models matter fluctuation computed at the current time ($z = 0$) with respect to $\Lambda$CDM as a function of model parameter ($\delta$ and $C$). We set $\Lambda$CDM in Planck 2018 results. In the upper left plot, we show the relative deviation for THDE model specifically, in the range of $1 \le \delta \le 3$, we obtain that the maximum relative difference occurs for $\delta = 1$ which means $w_{\rm de} \rightarrow 0$ according with Eq. (\ref{eos-thde}). The relative difference decreases as the value of $\delta$ increases, and from $\delta \approx 1.6$, we have $\Delta \delta_{\rm m0} > 0$. For $\delta = 2$, we have $w_{\rm de} = -1$, independent of the value of density parameter $\Omega_{\rm de}$, in this way the cosmological constant is recovered. The clustered and non-clustered dark energy have a similar behavior. In the case of $\delta > 2$, $w_{\rm de} < -1$, i.e., EoS parameter cross phantom barrier, thus the relative deviation increases. By considering the THDE2 (upper right), we assume $B = 3$ and, the interval $0.4 \le \delta \le 1.4$. As the THDE, clustered and non-clustered THDE2 evolves similarly. For $\delta = 0.4$, the relative difference is maximum. The cosmological constant is recovered in $\delta \approx 1$ and this moment, the difference is $\approx 0$. Thereafter, the relative difference increases $\Delta \delta_{\rm m0} > 0$, which means that matter perturbation in the THDE2 model computed in actual epoch is greater than by $\Lambda$CDM. In limit of $ \delta \rightarrow 1.4 $, EoS parameter associated with THDE2 model crosses phantom barrier. Therefore, we conclude that today matter perturbations is amplified. Finally, by regarding the HDE model we vary $C$ in the range $0.4 \le \delta \le 2$ and we find that the  $\Delta \delta_{\rm m0}$ is a decreasing function of $C$ and it belongs in the range $\sim[-18 \%, 10 \%]$. For $C \approx 0.83$, we have the line that divides the phantom from the quintessence region. In the case of small values of $C$, the phantom barrier is crossed thus, the corresponding matter perturbations in clustered and non-clustered DE are larger concerning $\Lambda$CDM. In the sense of higher values of $C$, we obtain that the matter perturbations are fewer in the HDE. We get that the $\Delta \delta_{\rm m0}$ between clustered and non-clustered HDE are significant when $C > 0.6$.

Furthermore, we study the THDE perturbations computed at the current time ($z = 0$) as a function of the model parameter. In Fig. \ref{fig:delta_de}, we show the three models studied in this work. In the upper left, we have the $\delta_{\rm de0}$ for THDE in the range of $1 \le \delta \le 3$. The maximum value of the perturbation occurs for $\delta = 1$, which means there is a clustering of THDE. We can note that the influence of this value appears in matter perturbations, as shown in Fig. \ref{fig:Delta_d_m} (upper left). In the case of $\delta = 2$, we have  $\delta_{\rm de0} \approx 0.001$ that is approximately the cosmological constant perturbation ($\delta_{\rm de0} = 0$). After this value, the THDE starts to behave like phantom dark energy, and the value of perturbation becomes negative. This value means that the dark energy density tends to form a void \cite{Maor2005,Abramo2007}. Now, considering the THDE2 model (upper right), note that value of $\delta_{\rm de0}$ oscillates between $\sim [0.005, -0.010]$ with the  maximum value in $\delta \approx 0.68$, $\delta_{\rm de0} \approx 0.026$. After the maximum value, the perturbation becomes in a decreasing function of $\delta$. Lastly, the HDE model is shown in lower center. Note that, the current THDE perturbation increases (see the term $1+w_{\rm de,i}$ in Eq. (\ref{eq:ini})) as a function of $C$. For $C  \approx 0.83$ we have $\delta_{\rm de0} \approx 0.009$.

In the following, we analyze the impact of different parameter combinations on the linear growth rate, Eq. (\ref{eq:linear-growth-rate}). In Fig. \ref{fig:growth-rate} we show the evolution of the linear growth function for holographic dark energy models studied in this work. The line fixed in $f = 1$ means a Einstein-de Sitter Universe ($\Omega_{\rm m} = 1$). The linear growth rate for the $\Lambda$CDM model turns to a constant value for high redshift since DE is very subdominant. In this case $f \rightarrow 1$.  In the THDE model (upper left), $f$ varies in intermediate redshift. Still, in the case of values close to $\delta = 2$, DE perturbations tend to compensate for the change in background evolution due to THDE. Note that, for the value close to $\delta = 1$ (red line) the matter perturbations take time to stabilize in $f \rightarrow 1$. Equations (\ref{eq:sec-ord-delta_m}) and (\ref{eq:sec-ord-delta_d}) capture its behavior.
The THDE perturbations improve matter clustering via the term ($\Omega_{\rm m}\delta_{\rm m} + \Omega_{\rm de}\delta_{\rm de}$) in Eq. (\ref{eq:sec-ord-delta_m}). Since the THDE model has a slightly decelerated background expansion than $\Lambda$CDM during the matter era, which in turn makes dark matter clustering less efficient, the contribution of THDE perturbations tends to atone this change, making dark matter growth more similar to the $\Lambda$CDM at high $z$. In the THDE2 (upper right), we obtain that some parameters combinations the growth rate in the late-time is higher than $\Lambda$CDM. The evolution of $f$ in the case of ($\Omega_{\rm m}$, $\delta$, $B$) = ($0.22$, $0.87$, $3.50$) is more slower than the others parameters combinations and, it converges for $f \rightarrow 0.9$ in high redshift. Finally, Fig. \ref{fig:growth-rate} (lower center) shows the evolution of the linear growth rate for the HDE model. Note that, in low redshift, the non-clustered HDE is similar to $\Lambda$CDM, but in $z \sim 0.8$, the evolution deviates and follows in a fewer value than the standard model. In the case of clustered HDE, the evolution in low redshift is opposite to the previous case, but in $z \sim 1.8$, the evolution becomes analogous to non-clustered HDE.

\section{Observational data and methodology}\label{sec:data-results}

To constrain the free parameters and compare the models under consideration, we perform a Bayesian statistical analysis using recent cosmological probes, which will be summarized as follows:

\begin{itemize}
\item \textbf{Baryon Acoustic Oscillations (BAO)}: We consider the set of $14$ BAO final measurements from Sloan Digital Sky Survey (SDDS) covering $8$ diverse redshift range. We use the measurements compiled in Table $3$ in \cite{Alam:2020sor}, regarding only BAO data.	
\item \textbf{BBN}: Measurement on the baryons from Big Bang nucleosynthesis as $100\Omega_{\rm b}h ^2 = 2.235 \pm 0.016$ \cite{Cooke2018}.

\item \textbf{CMB priors}: We assume the compressed CMB likelihood on the shift parameter $\mathcal{R}$, the acoustic scale $\ell_A$, and the baryon density $\Omega_b h^2$ \cite{aghanim2018,chen2019}. We fixed spectral index $n_s$ in the best-fit value and marginalized the CMB likelihood over $n_s$ by excluding the corresponding rows and columns from the covariance matrix.

\item \textbf{Cosmic Chronometers (CC)}:  The cosmic chronometers, obtained through differential age method, will be used here, i.e., the $31$ measurements of the Hubble parameter in the redshift range $0.07 < z < 1.965$ listed in \cite{GmezValent2018}.

\item \textbf{Type Ia Supernovae (SNe Ia)}: Supernovae are considered standard candles and are a powerful probe of cosmology, especially of the EoS of dark energy. The Pantheon sample is the most recent SNe Ia sample, consisting of $1048$ measurements in the redshift range $0.01 < z < 2.3$ \cite{scolnic2018}.

\item \textbf{Redshift-Space Distortion (RSD)}: To make a complete present work on the contribution of HDE models in the matter perturbations evolution, we use $f\sigma_8$ measurements from redshift-space distortion observations according to the Gold-$2017$ compilation. This compilation comprises the redshift range $0.02 < z < 1.52$. We use the formalism describes in Refs. \cite{Nesseris2017, Quelle2020}.

\end{itemize}

The usual approach to perform the Bayesian inference is to assume the prior distribution for the free parameters and the data distribution (likelihood). For Bayesian model comparison analysis, we will follow the standard steps describe in Refs. \cite{santos2017, santos2017b, santos2018,cid2019,andrade2018,dasilva2019,Silva2019}. For analysis of the Bayes factor, we use Jeffreys's scale reported in  \cite{Trotta2008}. In our analysis, we consider the $\Lambda$CDM as the standard model.  Moreover, as mentioned in Sect. \ref{sec:THDE}, we named the THDE with Hubble horizon,  THDE with future event horizon as THDE2, and standard HDE. For the Bayes factor $\ln \mathcal{B} < 0 $ we have support to $\Lambda$CDM model.

Using the above mentioned cosmological observations, we adopt the Nested Sampling \cite{skilling2004} method based on a Monte Carlo technique targeted at the efficient calculation of the evidence, yet which allows posterior probability as a by-product. Thus, by using the public package \textsf{MultiNest} \cite{feroz2007, Feroz2019, buchner2014} through the \textsf{PyMultiNest} interface \cite{pymultinest}. We follow $3$ steps: \textit{(i)} we run the codes with background data (BAO + BBN + CMB priors + CC + SNe Ia); \textit{(ii)} we consider the clustered DE and run the codes using background data + RSD; \textit{(iii)} we run the codes considering the non-clustered DE using background data + RSD. In order to perform the analysis, let us use uniform priors about the free parameters of the models investigated. The priors are $H_0 \sim \mathcal{U}(55.84, 90.64)$, $\Omega_{\rm m} \sim \mathcal{U}(0.001, 0.99)$,  $\delta \sim \mathcal{U}(1, 3)$ for THDE model, $\delta \sim \mathcal{U}(0.4, 1.4)$ for THDE2 model, $B \sim \mathcal{U}(0.0, 6.0)$, $C \sim \mathcal{U}(0.4, 2.0)$ and, $\sigma_8 \sim \mathcal{U}(0.4, 2.0)$. We fix the radiation density parameter in $\Omega_{\text{r}} h^2 = 1.698 \Omega_\gamma$ with $\Omega_\gamma = 2.469 \times 10^{-5}h^2$ and the baryon density in $\Omega_{\rm b}h^2 = 0.0224$ where $H_0 = 100 h \ \text{km} \ \text{s}^{-1} \text{Mpc}^{-1}$.

\section{Results and discussion}\label{sec:results}
In the course of this section, we will present our main results obtained from the statistical analysis using observational data from various cosmological probes, assuming the scenarios defined in Sect. \ref{sec:THDE}. The results are compiled in Table \ref{tab:results1} when we show the means and $1\sigma$ uncertainties of the free parameters considering the data combination. In Tables \ref{tab:results2} and \ref{tab:results3}, we show the results considering the background and RSD data for non-clustered and clustered DE, respectively. At the same time, to compare the HDE models with the $\Lambda$CDM, we present in Fig. \ref{fig:confidence_regions}, the corresponding contour plots for each model. In Fig. \ref{fig:wz}, we show the evolution of EoS parameter for all models considering the statistical results. Additionally, we will implement the model selection between HDE models and $\Lambda$CDM. 

\subsection{Parameter constraints}

\begin{figure*}[t]
	\centering
	\includegraphics[width=8.5cm]{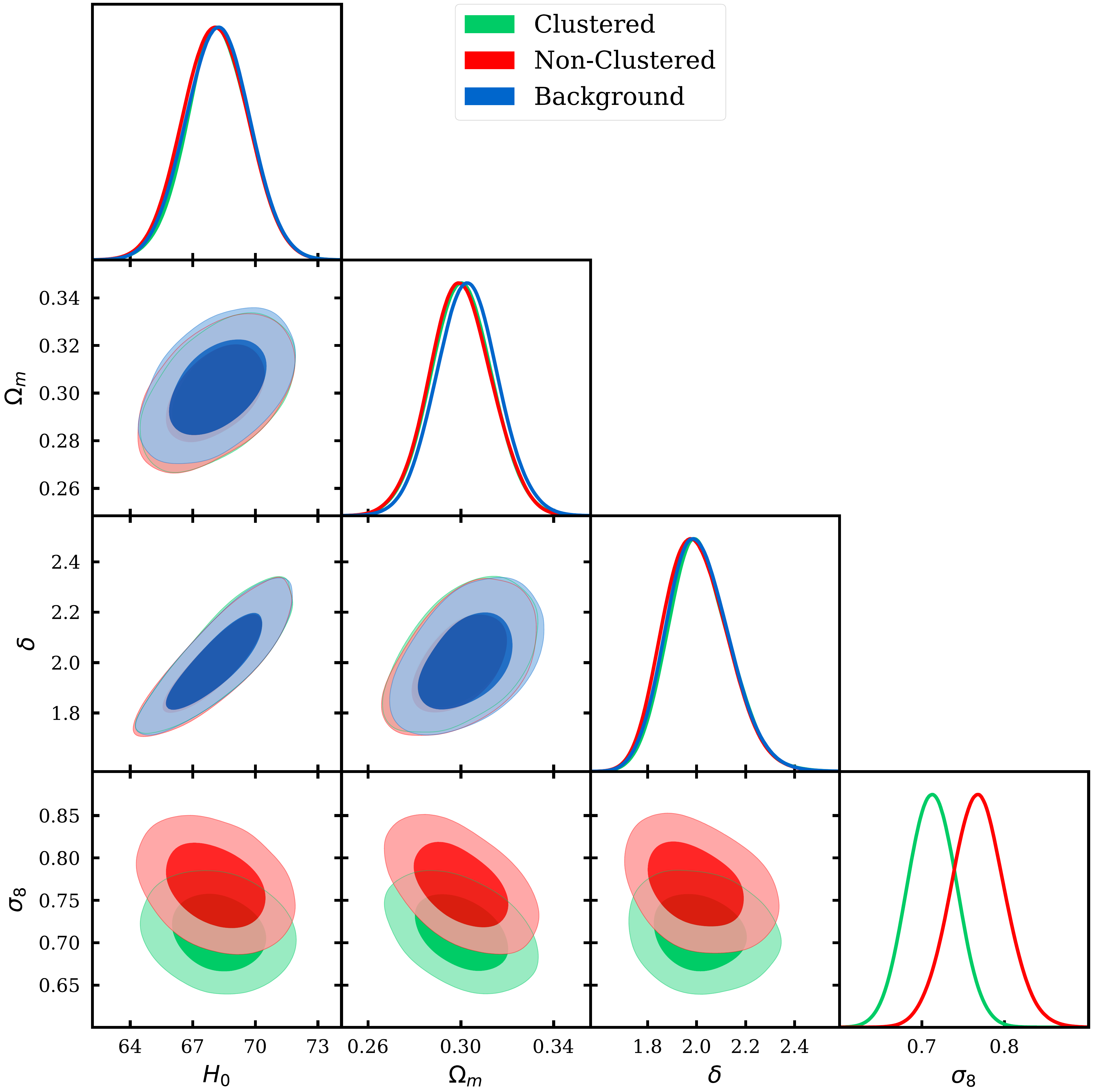}
	\includegraphics[width=8.5cm]{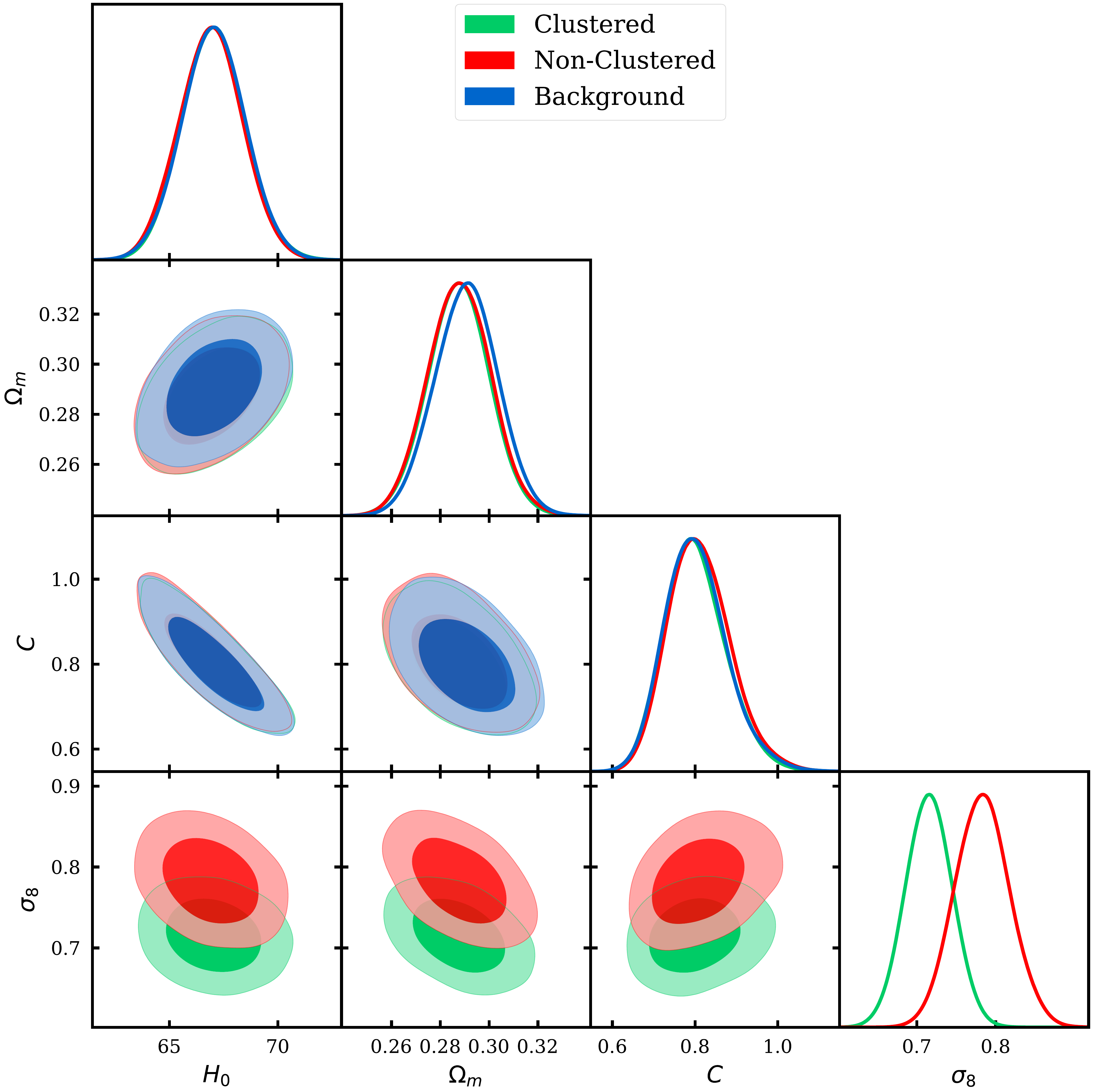}
	\includegraphics[width=9.5cm]{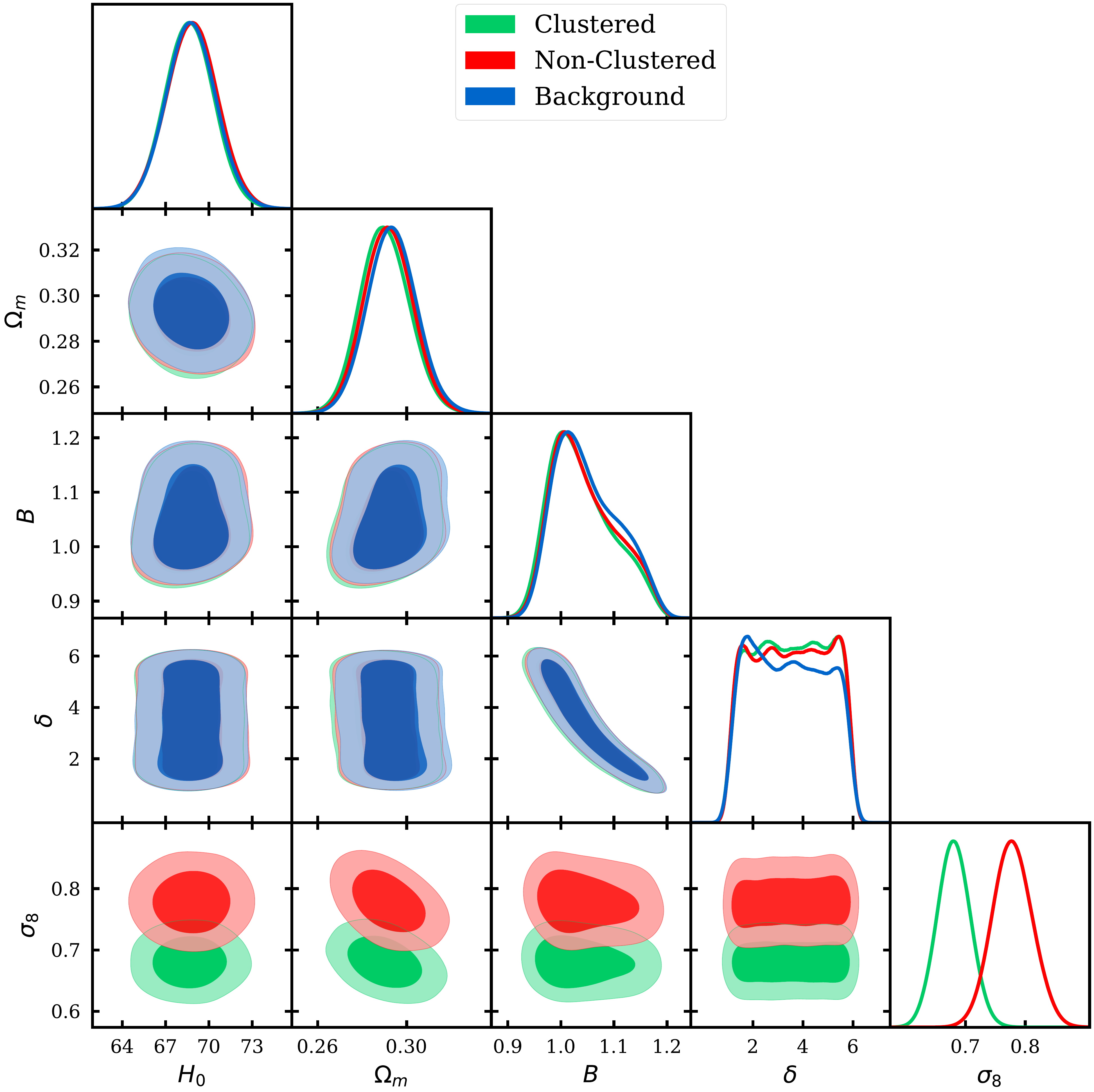}
	\caption{\label{fig:confidence_regions}$1\sigma$, $2\sigma$ and $3\sigma$ confidence regions for THDE (upper left), THDE2 (upper right) and HDE (lower center) models using several data combinations. The background data (blue contours) means BAO + BBN + CMB + CC + SNe Ia, Clustered (green contours) is the background + RSD data using the clustered DE and, non-clustered (red contours) is the same previous data combination but using non-clustered DE. B in units where $M^{2}_{P\ell} = 1$}.
\end{figure*}

Here, let us show the results obtained for HDE models considering background data. These results are summarized in the Table \ref{tab:results1} and the blue contours in Fig. \ref{fig:confidence_regions}. Our statistical analyses show that for THDE, the value obtained for Hubble constant $H_0 = 68.2 \pm 1.5 \ \text{km} \ \text{s}^{-1} \text{Mpc}^{-1}$, it is close to one obtained by the $\Lambda$CDM. The matter density value achieved is consistent in $1\sigma$ C.L with $\Lambda$CDM. Concerning the new physical parameter, we obtained $\delta = 2.01^{+0.11}_{-0.14}$ a slightly deviation from $\delta = 2$ (cosmological constant), however this value is included in $1\sigma$. The results for THDE model are in concordance with the Ref. \cite{Sadri2019}. Regarding THDE2 model, we obtained that the values for Hubble constant and matter density are compatible in $1\sigma$ C.L. with $\Lambda$CDM. Now, considering the nonadditive parameter and $B$, we obtain $\delta = 1.055^{+0.049}_{-0.078}$ and $B = 3.3 \pm 1.5$, respectively. The results for THDE2 show that there is no significant departure from the HDE model ($\delta = 1$ and $B = 3$). In fact, the two scenarios are in concordance with $1\sigma$ confidence, and furthermore the results obtained in this work are in agreement to related in Refs. \cite{Saridakis2018,DAgostino2019,Sadri2019}. The results for the standard holographic dark energy parameter $C$ are in agreement with the results obtained recently \cite{Wang2016,Akhlaghi2018,Zhang2020}. Concerning the tension of $H_0$, our results slightly alleviate this issue in $2.822\sigma$ for THDE, $3.425\sigma$ for THDE2, and $3.403\sigma$ for HDE.
\begin{table}[H]
	\centering
	\caption{\label{tab:results1} Confidence limits associated with the cosmological parameters through the background data. The columns show the constraints on each model whereas the rows show the parameter considering in this analysis. B in units where $M^{2}_{P\ell} = 1$}.
	\begin{tabular} {ccccc}
		\hline
		Parameter  & THDE & THDE2 & HDE &  $\Lambda$CDM \\
		\hline
		{\boldmath$H_0$}         & $68.2\pm 1.5 $ & $67.2\pm 1.4 $ & $67.0\pm 1.5$ & $68.22 \pm 0.82  $ \\
		
		{\boldmath$\Omega_m$}    & $0.303\pm 0.013 $ & $0.293\pm 0.011 $ & $0.291\pm 0.013  $ & $0.303\pm 0.012 $\\
		
		{\boldmath$\delta$}      & $2.01^{+0.11}_{-0.14}$ & $1.055^{+0.049}_{-0.078}$ & - & - \\
		
		{\boldmath$B$}           & - & $3.3\pm 1.5   $ & - & - \\
		
		{\boldmath$C$}           & - & - &$ 0.803^{+0.064}_{-0.082}  $ & - \\
		\hline
	\end{tabular}
\end{table}

Considering the combination between background and RSD data, we achieved the results shown in Table \ref{tab:results2} and the red contours in Fig. \ref{fig:confidence_regions} for the case of non-clustered DE. Note that $H_0$, $\Omega_{\rm m}$, $\delta$, $B$ and $C$ were slightly affected even though adding perturbations and RSD data. From Table \ref{tab:results2} we can read $\sigma_8 = 0.767\pm 0.032$ for the  $\Lambda$CDM  model, whereas the THDE prediction is $\sigma_8 = 0.768\pm 0.033$, THDE2 is $\sigma_8 = 0.781\pm 0.034$, and HDE is $\sigma_8 = 0.783\pm 0.035$. The values obtained for HDE models were similar to $\Lambda$CDM. Notice our results indicate that the tension of $\sigma_8$ considering THDE model is $1.282\sigma$, THDE2 model is $0.868\sigma$, and HDE is $0.788\sigma$. We conclude that non-clustered HDE models could alleviate the tension of $\sigma_8$. 

\begin{table}[H]
	\centering
	\caption{\label{tab:results2} Confidence limits associated with the cosmological parameters using the background + RSD data and non-clustered DE. The constraints on each model are shown in the columns, whereas the rows show the parameter considering in this analysis.}
	\begin{tabular} {lcccc}
		\hline
		Parameter & THDE & THDE2 & HDE &  $\Lambda$CDM \\
		\hline
		{\boldmath$h              $}  & $68.1\pm 1.5  $ & $67.1\pm 1.5    $ & $67.0\pm 1.5    $ & $68.19\pm 0.82     $ \\
		
		{\boldmath$\Omega_{\rm m} $}  & $0.300\pm 0.013 $ & $0.289\pm 0.013 $ & $0.288\pm 0.013    $ & $0.301\pm 0.012  $\\
		
		{\boldmath$\delta         $}  & $2.0^{+0.12}_{-0.14}$ & $1.055^{+0.049}_{-0.076} $ & - & - \\
		
		{\boldmath$B              $}  & - & $3.3^{+1.2}_{-2.1}  $ & - & -\\
		
		{\boldmath$C              $}  & - & - & $0.811^{+0.066}_{-0.082}   $ & - \\
		{\boldmath$\sigma_8       $}  &  $0.768\pm 0.033   $ &  $0.781\pm 0.034  $ & $0.783\pm 0.035$ & $0.767\pm 0.032    $ \\
		\hline
	\end{tabular}
\end{table}

Now, assuming the THDE can be clustered, we obtained the results which are provided in Table \ref{tab:results3} and the green contours in Fig. \ref{fig:confidence_regions}. In this case, the parameters $H_0$, $\Omega_{\rm m}$, $\delta$, $B$ and $C$ were a little deviated when adding RSD data. Concerning the $\sigma_8$, the results achieved were $ 7.3 \% $, $12.9 \%$ and $8.589 \%$ lower for THDE, THDE2 and HDE, respectively. This results for clustered THDE, the tension of $\sigma_8$ increased for $3.235\sigma$, $4.574\sigma$ and $3.137\sigma$.
\begin{table}[H]
	\centering
	\caption{\label{tab:results3} Confidence limits for the cosmological parameters using the background + RSD data and clustered DE. The columns show the constraints on each model whereas the rows show the parameter considering in this analysis.}
	\begin{tabular} {lccc}
	\hline
	Parameter & THDE & THDE2 & HDE  \\
	\hline
	{\boldmath$h              $}  & $68.2\pm 1.5  $ & $67.1\pm 1.7   $ & $67.0\pm 1.5      $ \\
	
	{\boldmath$\Omega_{\rm m} $}   & $0.300\pm 0.013  $ & $0.290\pm 0.011$ & $0.288\pm 0.013     $ \\
	
	{\boldmath$\delta         $}  & $2.01^{+0.11}_{-0.14}  $ & $1.054^{+0.046}_{-0.076}$ & - \\
	
	{\boldmath$B              $} & - & $3.5^{+2.1}_{-1.3} $ & - \\
	
	{\boldmath$C              $} & - & - &$0.801^{+0.065}_{-0.081} $ \\
	{\boldmath$\sigma_8       $}  &  $0.712\pm 0.030 $ &  $0.680\pm 0.028   $ & $0.715\pm 0.030  $ \\
	\hline
\end{tabular}
\end{table}

\begin{figure}[H]
	\centering
	\includegraphics[width=8.5cm]{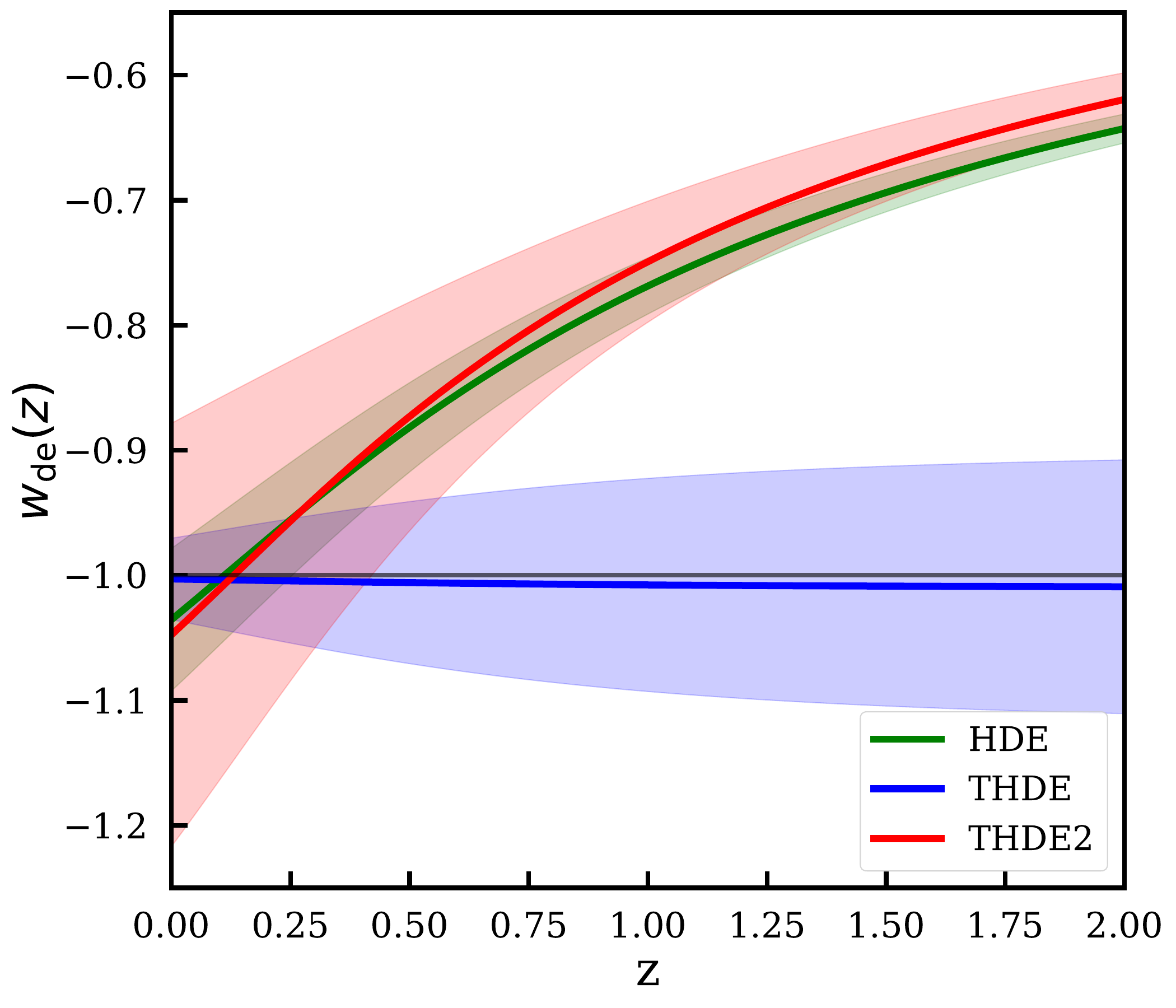}
	\caption{\label{fig:wz} The evolution of equation of state parameter of HDE models with respect to the redshift $z$ with means and $1\sigma$ errors of parameters. The continuous line is the mean and shadow region, error. The black line means $w_{\rm de} = -1$.  We use the propagation of uncertainties described in Ref. \cite{Xu2013} and the Python module Uncertainties \cite{Lebigot2016}.}
\end{figure}

Let us analyze the evolution of the equation of state parameter, given by Eqs. (\ref{eos-thde}), (\ref{eq:eos-thde2}) and (\ref{eq:conservation-hde}) assuming the means and errors values from the statistical analysis. As shown in Fig. \ref{fig:wz}, it is seen that the $w_{\rm de}$ can cross the phantom barrier, i.e., $w_{\rm de} < -1$. We get the current values of the equation of state parameter $w_{\rm de }(z = 0) = -1.003 \pm 0.033$, $-1.05 \pm 0.17 $ and $-1.040 \pm 0.060$ for THDE, THDE2 and HDE models, respectively. The phantom behavior of dark energy indicates that the energy density increases over time (instead of the constant cosmological model). If the Universe is accelerating due to phantom dark energy, it may end in several types of future singularities (see Ref. \cite{Ludwick:2017tox} for phantom dark energy review). Our results for THDE2 are different from those obtained in Ref. \cite {DAgostino2019}, which obtained the quintessence-like behavior, but the phantom-divide crossing line is not permitted. In fact, in this work, we consider BBN + BAO + CMB priors + RSD data in addition to SNe Ia and CC. In this way, we obtained slightly different results from Ref. \cite {DAgostino2019}.

\subsection{Model selection}

For the sake of the model comparison, we calculated the Bayes factor considering $\Lambda$CDM as the reference model. In order to obtain the Bayes factor values for each model, we consider the logarithm of the evidence obtained from the \textsf{MultiNest} and do the following operation $\ln \mathcal{B} = \ln\mathcal{E}_{\text{model}} - \ln\mathcal{E}_{\Lambda \text{CDM}}$. The Jeffreys scale interprets the Bayes factor as follows: inconclusive if $|\ln \mathcal{B}| < 1$, weak if $1 \leq |\ln \mathcal{B}| < 2.5$, moderate if $2.5 \leq |\ln \mathcal{B}| < 5$ and strong if $|\ln \mathcal{B}| \geq 5$. A negative (positive) value for the Bayes factor means that the competing model is disfavored (supported) to the $\Lambda$CDM model. In Table \ref{tab:bayes-factor}, we show the values obtained for the logarithm of evidence, the logarithm of Bayes factor, and the interpretation of Bayes factor from the Jeffreys scale.

Looking at the background data, the THDE model presents weak evidence against it, while others have moderate evidence against them. Regarding the background data with RSD in non-clustered DE, the conclusion is the same as the previous one. Finally, assuming that holographic dark energy can cluster and the background and RSD data, we obtain that model THDE2 has strong evidence against it and others, they have moderate evidence against them. We conclude that, from a Bayesian evidence point of view, the observational data discard the models studied in this work.

\begin{table}[H]
	\centering
	\caption{\label{tab:bayes-factor} The table shows data, model, logarithm of evidence,  logarithm of Bayes factor and interpretation. We use the interpreation given by Jeffreys scale \cite{Trotta2008}.}
	\begin{tabular}{clccc}
\hline
Data                                                                      & Model        & $\ln \mathcal{E}$ & $\ln \mathcal{B}$ & Interpretation     \\ \hline
\multirow{4}{*}{Background}                                               & $\Lambda$CDM & -543.413          & 0.0               & -                  \\
                                                                          & THDE         & -545.242          & -1.811            & Weak (Against)     \\
                                                                          & THDE2        & -547.176          & -3.763            & Moderate (Against) \\
                                                                          & HDE          & -546.237          & -2.824            & Moderate (Against) \\ \hline
\multirow{4}{*}{\shortstack{with RSD \\ Non-Clustered DE}} & $\Lambda$CDM & -552.324          & 0.0               & -                  \\
                                                                          & THDE         & -554.143          & -1.819            & Weak (Against)     \\
                                                                          & THDE2        & -556.254          & -3.819            & Moderate (Against) \\
                                                                          & HDE          & -554.991          & -2.667            & Moderate (Against) \\ \hline
\multirow{4}{*}{\shortstack{with RSD \\ Clustered DE}    }     & $\Lambda$CDM & -552.324          & 0.0               & -                  \\
                                                                          & THDE         & -555.332          & -3.008            & Moderate (Against) \\
                                                                          & THDE2        & -559.855          & -7.531            & Strong (Against)   \\
                                                                          & HDE          & -555.797          & -3.473            & Moderate (Against) \\ \hline 
\end{tabular}
\end{table}

\section{Conclusions}\label{sec:conclusions}

In this work, we studied three different HDE models considering both the perturbative level and observational viability. The models were (i) the THDE with Hubble horizon and (ii) future event horizon as cutoffs (THDE2), and (iii) Standard HDE with future event horizon as a cutoff. We implemented the linear perturbation theory and studied the growth of matter perturbations in clustered and non-clustered HDE. Moreover, we have performed a Bayesian analysis to compare models using the most recent observable data: (i) baryon acoustic oscillations measurements, (ii) big bang nucleosynthesis, (iii) cosmic microwave background priors, (iv) cosmic chronometers, v) Pantheon type Ia supernovae, and (vi) redshift space distortions data.

Firstly, we studied the relative deviation of matter fluctuation computed at the current time considering the HDE and $\Lambda$CDM model. We obtained the minimum difference in the values that holographic dark energy behaviors as the cosmological constant. Next, we analyzed the impact of different parameter combinations on the linear growth rate, and we found that clustered THDE models are always larger than non-clustered ones.

In the second moment, we analyzed the cosmological parameter behavior against the datasets considered. By assuming the priors for the free parameters described in Sect. \ref{sec:data-results}, we performed a complete Bayesian analysis on the joint likelihood of BAO + BBN + CMB + CC + SNe Ia. We found out that the THDE, with Hubble horizon cutoff, has a little deviation of the cosmological constant ($\delta = 2$) with equation of state parameter today $w_{\rm de}(z = 0) = -1.003 \pm 0.033$ which a crossing the phantom barrier. For THDE with future event horizon (THDE2), we obtained that there is no significant deviation from the standard HDE model with $w_{\rm de}(z = 0) = -1.05 \pm 0.17$. The models studied here can relieve $\approx 1\sigma$ the $\sigma_8$ tension in the non-clustered case. Considering the $H_0$ tension, these models can alleviate in $\approx 2.8\sigma$ this problem.

Finally, we compared the new HDE with $\Lambda$CDM using the Bayesian evidence for all data combinations. The analysis demonstrated that both, a variety of geometric data and growth data of structure, discard models of THDE. So, we could conclude that the dark energy models studied here cannot fit the observations in cluster scales as goodly as the $\Lambda$CDM model.

Despite the strong observational bounds obtained in this work, we cannot wholly exclude THDE. So it is expected that future experiments such as J-PAS \cite{Benitez2014}, DESI \cite{Aghamousa2016}, and Euclid \cite{Euclid2011} improve the quality of $f\sigma_8$ data, and thus will help to mitigate or exclude the tensions in cosmology.  

\begin{acknowledgments}
We  thank  the  anonymous  referee  for  the  comments that  helped  us  greatly  improve  this  paper. The authors thank Brazilian scientific and financial support federal agencies, Coordena\c c\~ao de Aperfei\c coamento de Pessoal de N\'{\i}vel Superior (CAPES) and Conselho Nacional de  Desenvolvimento  Cient\'{\i}fico  e  Tecnol\'ogico (CNPq).  RS thanks CNPq (Grant No. 307620/2019-0) for financial support. This work was supported by High-Performance Computing Center (NPAD)/UFRN.
\end{acknowledgments}

\bibliographystyle{apsrev4-2}
\bibliography{references}

\end{document}